\documentclass[letterpaper,twocolumn,10pt]{article}
\usepackage{usenix2019_v3}

\usepackage{tikz}
\usepackage{amsmath}
\usepackage{booktabs}
\usepackage{subcaption}
\usepackage{xurl}
\usepackage{pifont}
\usepackage{xcolor}

\usepackage{filecontents}
\usepackage{graphicx}

\begin{document}
%-------------------------------------------------------------------------------

\pagestyle{plain}

\date{}

\title{\Large \bf Achieving Cloud-Grade SLOs for Local Mixture-of-Experts Inference through CPU–GPU Hybrid Design\thanks{Accepted to the 20th USENIX Symposium on Operating Systems Design and Implementation (OSDI '26). The official version will appear in the OSDI '26 proceedings published by USENIX.}}

\author{
\begin{tabular}{c}
Wenxin Wang\textsuperscript{1},
Yule Hou\textsuperscript{2},
Yu Ji\textsuperscript{2},
Peng Qu\textsuperscript{1,3},
Youhui Zhang\textsuperscript{1,3}
\\[0.6em]
\textsuperscript{1}\emph{Tsinghua University}
\\
\textsuperscript{2}\emph{Xingyun Integrated Circuits Co., Ltd.}
\\
\textsuperscript{3}\emph{Beijing National Research Center for Information Science and Technology}
\end{tabular}
}

\maketitle
\thispagestyle{plain}
%-------------------------------------------------------------------------------
\begin{abstract}
%-------------------------------------------------------------------------------
Local deployment of large Mixture-of-Experts (MoE) models falls short of the service quality achieved in cloud-scale environments, even under low-concurrency workloads. We identify four key gaps in local MoE inference: reliance on capacity-reduced models (quantized, distilled, rerouted), inability to meet 30-second TTFT for long prefills (>12K), sub-baseline decode throughput (<20 tokens/s), and poor concurrency under mixed prefill–decode and batched decode workloads. We present a CPU–GPU hybrid system that achieves cloud-level SLOs on dual-socket commodity CPUs and consumer GPUs by (1) stream-loading prefill (SLP), boosting prefill throughput to 1,200 tokens/s and enabling 32K prompts within 30 seconds; (2) distributed SLP (DSLP) with SmallEP expert parallelism, reaching 1,800 tokens/s and 45K prompts in 30 seconds on two RTX 5090s; (3) intra-node prefill–decode disaggregation with zero-copy shared weights and a dual-batch attention–MoE overlap scheme, sustaining concurrency with <15\% latency increase and 50\% throughput gains; (4) an AVX-512–optimized FP8 GEMV kernel, enabling native CPU FP8 inference while delivering 4–5× lower CPU latency; and (5) fine-grained CPU parallelism that attains 28 tokens/s on INT4 DeepSeek-V3 and 21.5 tokens/s on intact FP8 V3. Evaluations show our system delivers cloud-level QoS for flagship MoE models on consumer CPU–GPU platforms, reshaping local deployment with intact, original-precision inference and enabling high-quality, cost-effective access without datacenter infrastructure.
\end{abstract}

\section{Introduction}

Large-scale Mixture-of-Experts (MoE) architectures have become a central design point for state-of-the-art language models, enabling their parameter counts approaching the trillion scale while sustaining practical training and inference costs through sparse expert activation~\cite{kimiteam2025kimik2openagentic, liu2024deepseek, guo2025deepseek, qwen3}. This sparsity reshapes the deployment landscape. On one end, massive cloud deployments—often serving tens of thousands of concurrent requests—benefit from the highly overlapped expert activation, which drives hardware utilization toward peak levels~\cite{deepep2025, zuo2025serving}. On the other end, local or low-concurrency deployments benefit from the sparse activation from small batch sizes, which significantly cuts down memory traffic, reducing the cost of accessing expert parameters and enabling efficient inference on resource-limited hardware~\cite{kamahori2024fiddler, chen2025ktransformers}. 

Significant engineering effort has gone into scaling MoE inference in large cloud environments, where service-level objectives are well established~\cite{deepseekV3-overview, wang2025step} and system designs have converged on robust tradeoffs among throughput, latency, and cost. 
But these methods won't apply to local scenarios, where the hardware landscape and workload characteristics differ substantially.
In contrast, local deployment of MoE models remains underexplored. A promising approach for local deployment is CPU–GPU hybrid inference, which alleviates the constraints of limited local VRAM by storing the majority of model parameters in larger and more affordable system memory. 

Prior systems follow the arithmetic-intensity principle: compute-heavy layers—such as attention and shared experts—are assigned to the GPU, whereas lower-intensity, parameter-heavy components—such as routed experts—are offloaded to the CPU. The state-of-the-art system, KTransformers~\cite{chen2025ktransformers}, demonstrates that careful CPU–GPU orchestration can yield superior performance. By adopting quantized model and altering routing, KTransformers achieves an acceptable 22 tokens/s decoding throughput on an INT4-quantized, rerouted DeepSeek-V3/R1 model.

However, compared to service quality achievable in cloud-scale systems, current local deployment systems still lag substantially. This gap manifests along several dimensions.
(1) \textbf{Model quality}. Local deployments typically rely on quantized, distilled, rerouted, or otherwise capacity-reduced models, imposed by tight VRAM and memory-bandwidth budgets. Consequently, users cannot access the full-precision, unmodified MoE models commonly available in cloud services, leading to a pronounced gap in model quality.
(2) \textbf{Prefill Latency (time-to-first-token, TTFT)}. TTFT is critical for emerging agent workloads and document QA or summarization. Cloud services target a 30-second TTFT~\cite{qin2025mooncake}. In contrast, prior local systems—despite using advanced matrix units in high-end CPUs—struggle to meet the 30-second target for 12K-token or longer prefills on the DeepSeek-V3 INT4 model~\cite{chen2025ktransformers}.
(3) \textbf{Decode Latency (time-per-output-token, TPOT; or tokens-per-second, TPS)}. Cloud systems typically deliver >20 tokens/s per-request decoding as a responsiveness baseline~\cite{deepseekV3-overview, wang2025step}. Prior local systems, even with quantized models, fall short (e.g., 16 tokens/s on DeepSeek-V3 INT4~\cite{chen2025ktransformers}), leading to noticeably reduced responsiveness relative to cloud deployments.
(4) \textbf{Concurrency}. Two gaps emerge under concurrency even considering the inherent low concurrency requirements in local deployments. First, prefilling and decoding are often mutually exclusive on local systems, producing either impaired TPOT for decoding under prefill-first scheduling or long TTFT under decode-first scheduling. Second, concurrent decode requests—even at small batch sizes—can substantially degrade per-request latency (TPOT), as expert activations effectively multiply with each additional request. 

Throughout the paper, we use these cloud-level targets as local QoS reference points. Our key insight is that local MoE inference should be co-designed around the resources available on commodity nodes: large CPU DRAM capacity and bandwidth, consumer-GPU compute, limited VRAM, and PCIe-class interconnects. Based on this principle, our system reorganizes prefill, decode, parallelism, and concurrency to close the four SLO gaps locally. Our contributions include:

\begin{itemize}
    \item \textbf{Stream-loading prefill.} We introduce a stream-loading prefill (SLP) execution model that overcomes consumer GPU VRAM limits, maximizes GPU utilization. SLP increases prefill throughput to up to \textbf{1,200 tokens/s} and supports prefilling 32K-token prompts within a 30-second TTFT budget.
    \item \textbf{Distributed stream-loading prefill.}
    Building on the SLP model, we develop a parallelism strategy that enables multi-GPU prefilling in local settings, with a SmallEP optimization for expert parallelism tailored to regimes with small expert-parallel sizes. The distributed stream-loaded prefill (DSLP) design delivers over \textbf{1,800 tokens/s} prefill throughput and supports prefilling \textbf{45K-token} prompts within a 30-second TTFT budget using only two consumer-grade RTX 5090 GPUs—offering better cost efficiency than CPU-prefill approaches relying on high-end CPU matrix units.
    
    \item \textbf{Concurrency via intra-node prefill–decode disaggregation and dual-batch attention–MoE overlap.} For prefill–decode concurrency, we perform prefill–decode disaggregation within a node atop the SLP execution model, and employ zero-copy, shared model weights, enabling concurrent prefill and decode with minimal interference—less than 15\% latency increase. For batched decode concurrency, we introduce a dual-batch attention–MoE parallelism scheme that interleaves attention and MoE computations across two microbatches, increasing system throughput by 50\% with limited impact on per-sequence QoS.
    \item \textbf{High-performance FP8 GEMV kernel.} We develop an AVX-512–optimized FP8 GEMV kernel tailored for native FP8 MoE inference, achieving 4–5× lower latency and higher bandwidth utilization compared to existing BLAS libraries. This enables practical original-precision inference on CPUs.
    \item \textbf{Fine-grained CPU Parallelism for MoE Inference} We design fine-grained synchronizations, minimize inter-socket communication and synchronization overheads, and fuse quantization-specific operations on CPUs. This improves decode latency for both full-precision and quantized inference, delivering a 28 tokens/second decode rate on INT4 DeepSeek-R1 model (1.25× improvement over KTransformers) and 21.5 tokens/second on the intact FP8 DeepSeek-R1 model.
\end{itemize}

We evaluate our system on a dual-socket low-end server-class CPU platform with 1--2 consumer GPUs, against mainstream CPU inference frameworks (KTransformers, llama.cpp, ik\_llama.cpp) and state-of-the-art CPU BLAS libraries (OpenBLAS, AOCL-BLAS). Our results demonstrate that such consumer-grade CPU--GPU platforms can serve flagship large MoE models at cloud-level quality of service, at a fraction of the cost of GPU-centric supernode or cluster deployments. We further discuss the expected behavior on lower-end hardware, including single-socket CPUs, different CPU SKUs, and lower-VRAM GPUs in Section~\ref{sec:discussion}.

\section{Background and Motivation}

\subsection{Background}

\begin{figure}
    \centering
    \includegraphics[width=\linewidth]{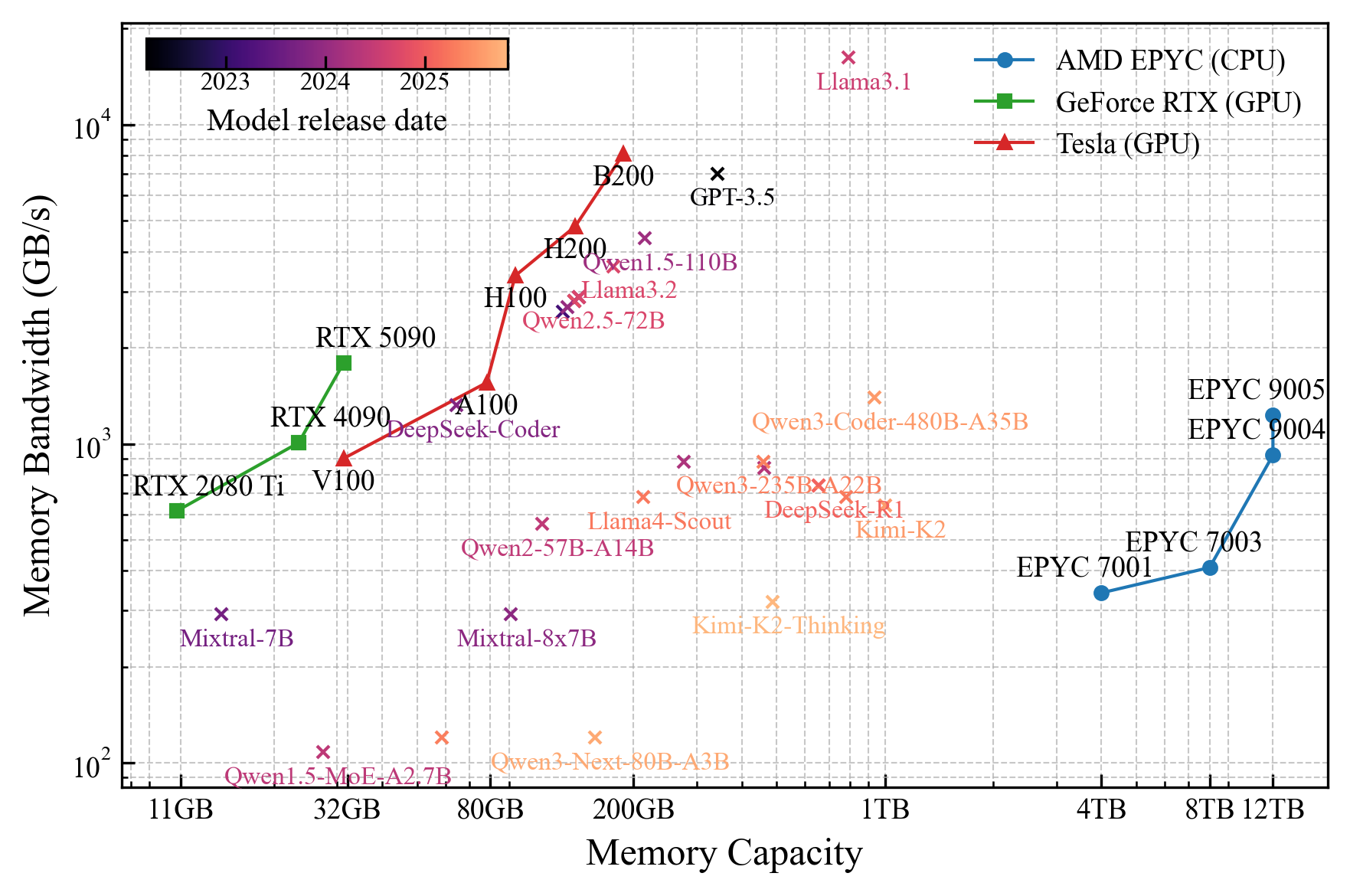}
    \caption{Models’ memory demands and device comparison. Model memory capacity reflects parameter size at original weight dtype; memory bandwidth reflects activated weight accesses under a 20 tokens/second service-level objective (i.e., 20 serial activated-weight reads per second). Devices shown include NVIDIA GeForce RTX, NVIDIA Tesla data center GPUs, and AMD EPYC CPUs.}
    \label{fig:model-device-cap-bw}
\end{figure}

Mixture-of-Experts (MoE) architectures introduce conditional computation into transformer models by replacing dense feed-forward blocks with large pools of expert networks\cite{shazeer2017outrageously}. A learned gating mechanism selects a small subset of experts—typically top-k or grouped top-k—for each token, enabling the model to scale its total parameter count without proportionally increasing compute per token. Modern designs such as DeepSeek\cite{liu2024deepseek, deepseek-ai_deepseek-v2_2024} series augment this with additional shared experts that are always active, providing a common representation path aside token-specific specialization.

To sustain inference throughput for models with large parameter counts, serving systems must deliver both high memory capacity and high effective memory bandwidth. Figure~\ref{fig:model-device-cap-bw} surveys recent models and situates their capacity and bandwidth requirements against contemporary device trends. With brighter colors indicating more recent releases, a clear trajectory emerges: total parameter capacity continues to climb (approaching 1 trillion parameters, or roughly 1 TB in raw weight size), while activated weights per token are reduced via selective activation. This shift lowers the bandwidth pressure for inference to approximately 1 TB/s, as evidenced by recent large MoE deployments\cite{kimiteam2025kimik2openagentic, liu2024deepseek}.

As Figure~\ref{fig:model-device-cap-bw} illustrates, modern multi-socket CPU systems can deliver aggregate DDR5 bandwidth in the same ballpark as prior GDDR or HBM solutions for the activated-weight access patterns typical of large MoE inference. Combined with substantially lower memory cost (about \$3 per GB for DDR5 versus about \$30 per GB for HBM) and improved system-level power efficiency, CPU-based deployments become particularly attractive for low-concurrency inference scenarios, where capacity per node and cost/performance at moderate bandwidth are the dominant considerations.

Leveraging modern CPU platforms, hybrid systems such as KTransformers\cite{chen2025ktransformers} report approximately 16 tokens/second for decode and up to 500 tokens/second for prefill on the DeepSeek-V3 Int4 model—yielding around 20 seconds TTFT for an 8K prompt and an estimated 30+ seconds for 12K or longer. Nonetheless, a substantial gap remains between local inference performance and cloud-level service quality. In the next section, we dissect this gap and its root causes.

\subsection{Motivations}

Below, we dissect the gap between state-of-the-art local deployment systems and cloud-level service quality across five dimensions—decode, prefill, multi-GPU parallelism, concurrency, and FP8 kernel support—and, based on this analysis, introduce targeted improvements for each. These challenges are not isolated. In local deployment, decode is limited mainly by CPU-side bandwidth efficiency, prefill by how much compute can be exposed under tight VRAM budgets, parallelism by commodity PCIe interconnects, and concurrency by whether these resource pressures can be separated and reorganized rather than forced onto a fixed execution path. This coupling motivates the key principle of our design: local MoE inference should be co-designed around CPU DRAM capacity/bandwidth and consumer-GPU compute, rather than directly inheriting cloud-style assumptions.

\vspace{-0.7em}
\subsubsection{Decode} 

To accommodate limited VRAM in local deployments, hybrid inference systems employ computation offloading\cite{chen2025ktransformers, kamahori2024fiddler, llamacpp, ikawrakow_ik_llama_cpp}: MoE layer weights are persisted in CPU memory and their expert computations execute on the CPU, while densely activated weights and corresponding kernels remain resident and execute on the GPU. Building on this split, KTransformers\cite{chen2025ktransformers} specifically targets decode-phase bottlenecks—CPU–GPU synchronization and kernel launch overheads—achieving state-of-the-art decode throughput compared to prior local systems such as llama.cpp and Fiddler. 

Despite these improvements, decode performance remains short of cloud-level responsiveness: KTransformers delivers approximately 16 tokens/second on the Int4 quantized model, whereas cloud services running full precision models commonly target $\geq$20 tokens/second as a responsive baseline\cite{deepseekV3-overview, wang2025step}. Decode in local deployment of large MoE models is predominantly memory-bandwidth bound. Using the published figures—74\% CPU time and 20.4B activated MoE parameters—KTransformers’ decode throughput implies an effective memory bandwidth of roughly 221 GB/s. On the reported dual-socket Xeon DDR5 system, this is only about 50\% of the nominal aggregate DDR5 bandwidth. This breakdown highlights significant hardware underutilization and motivates our work to close the gap to cloud-level service quality, as elaborated in Section~\ref{sec:cpu-backend}.

\vspace{-0.7em}
\subsubsection{Prefill}

In popular hybrid systems, sparsely activated MoE computations run on the CPU while densely activated components such as attention execute on the GPU. During prefill—where compute intensity rises—computation increasingly dominates end-to-end latency. As sequence length grows beyond roughly 512 tokens, CPU-side MoE execution saturates available compute, and its latency becomes the primary contributor to time-to-first-token (TTFT).

To mitigate CPU limitation, KTransformers leverages Intel’s Advanced Matrix Extensions (AMX) instruction set with carefully specialized CPU GEMM kernels, achieving state-of-the-art peak prefill throughput of over 500 tokens/second at a 2048-token prompt, surpassing prior hybrid approaches. Nonetheless, the approach remains bounded by CPU arithmetic throughput: CPU-centric pipelines cannot match GPU-resident dense GEMM performance. As prompt lengths grow to 8K, 16K and 32K (common in agent, document Q\&A and text summary tasks\cite{bai2024longbench, bai2024longbench2}), CPU-side MoE computation becomes the dominant bottleneck, driving time-to-first-token upward to around 20, 40 (estimated) and 100 (estimated) seconds.

Conversely, to exploit GPU compute, weight-offloading systems\cite{alizadeh2024llm, he2024expertflow, hwang2024pre, song2024promoe, tang2024hobbit} prefetch model weights to the GPU prior to kernel launches, using predictive scheduling, quantized or compressed weights and reordered pipelines to reduce reactive latency. These designs typically (1) apply a unified optimization strategy to both decode and prefill that prioritizes minimal transfer latency and proactive fetching, and (2) target edge deployments of small- to mid-scale models with lower compute intensity. Consequently, they are optimized for latency rather than throughput, making them ill-suited for compute-heavy prefill on large MoE models and deliver suboptimal prefill performance compared to contemporary computation-offloading approaches.

These limitations motivate a different direction: keep the benefits of weight offloading to avoid CPU-bound compute, but focus explicitly on prefill and optimize for throughput on large MoE models across both weight transfer and computation. In Section~\ref{sec:prefill} we introduce a fine-grained stream-loading prefill (SLP) mechanism that pipelines and fully overlaps expert weight transfer with GPU kernel execution at sub-layer granularity. SLP eliminates reactive stalls without prediction overhead, quantization, or rerouting—preserving model quality while sustaining low TTFT at scales.

\vspace{-0.7em}
\subsubsection{Multi-GPU Parallelism}
\label{sec:bg:parallelism}

\begin{figure}
    \centering
    \includegraphics[width=0.95\linewidth]{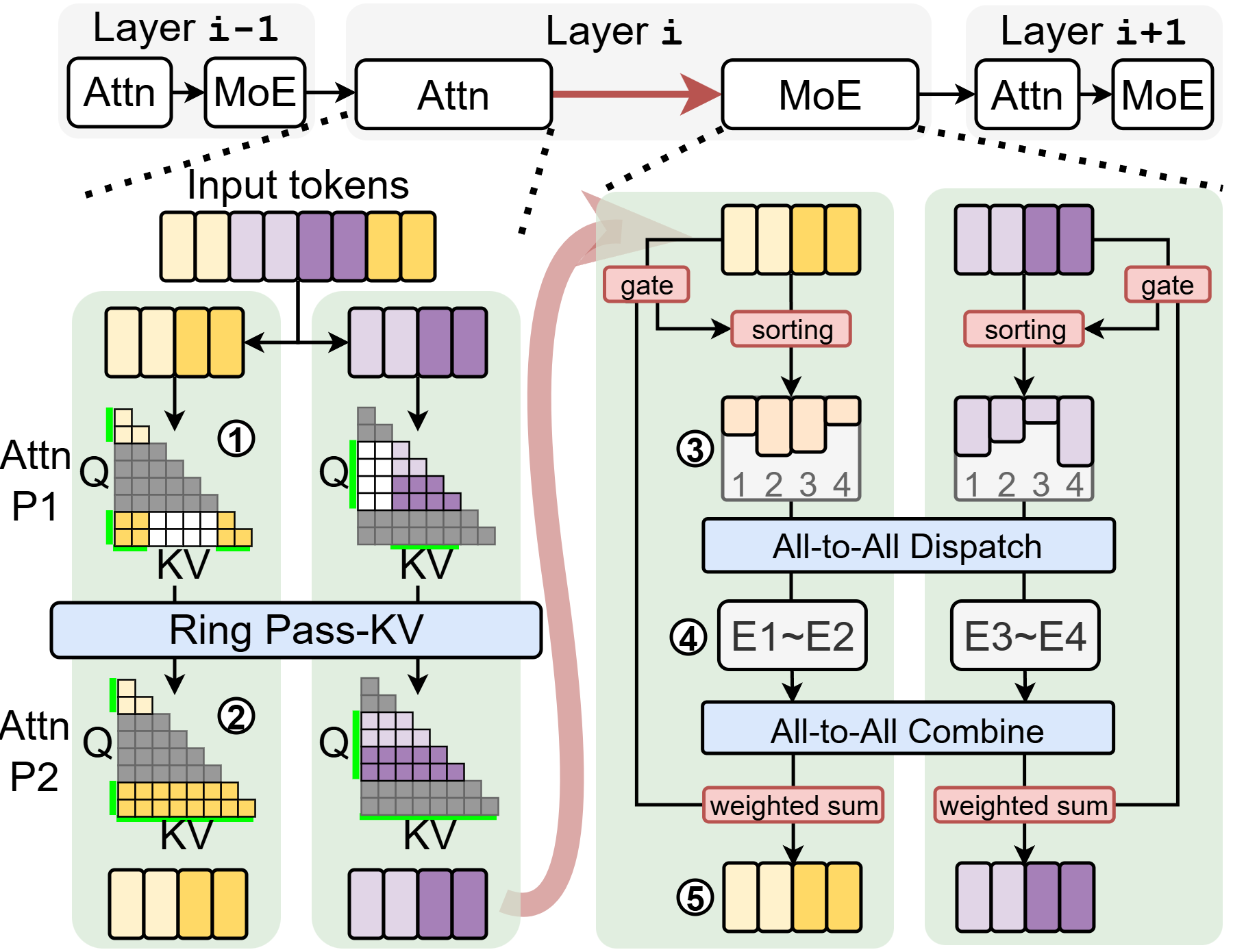}
    \caption{Example workflow: zigzag StripedAttention (Pass-KV, CP=2, left) followed by expert parallelism (EP=2, right). Tokens are split across two ranks; each rank attends locally (phase 1), performs one ring Pass-KV exchange, then completes attention (phase 2). Outputs then flow through MoE gating, dispatch/compute/combine, and weighted aggregation.}
    \label{fig:spep}
\end{figure}

To accelerate prefill, large-scale GPU deployments rely on context, sequence, tensor, pipeline, and expert parallelism\cite{yang2024context, liu2023ring, brandon2023striped, jacobs2023ulysses, shoeybi2019megatron, zhu2024nanoflow, deepep2025, qin2025mooncake, lepikhin2020gshard}. Local deployments, however, operate at a different point: low concurrency with strict latency targets, PCIe-only interconnects often lacking GPU–GPU P2P, and constrained VRAM. Under these constraints, many cluster-oriented schemes do not translate directly. We implement a communication-lean hybrid of context and expert parallelism to distribute long-context prefill; yet, even with reduced exchanges, PCIe dispatch/combine remain significant, leaving communication as the primary bottleneck.

Figure~\ref{fig:spep} illustrates StripedAttention (CP=2, Pass-KV)\cite{brandon2023striped, zhuzilin_ring_flash_attention} combined with standard expert parallelism (EP=2)\cite{lepikhin2020gshard, deepep2025}. StripedAttention splits input tokens evenly across two ranks. Each rank first computes causal attention over its local KVs (Attn P1, \ding{172}), then performs a ring Pass-KV to exchange the remaining KVs, with which each rank completes the causal attention (Attn P2, \ding{173}). In the expert-parallel stage, tokens on each rank are gated and sorted to form per-expert token lists (\ding{174}). An All-to-All Dispatch exchanges tokens so each GPU receives the tokens assigned to its local experts. Local experts execute on those tokens (\ding{175}), after which an All-to-All Combine returns the expert outputs for the originating tokens. A final weighted sum produces the per-token outputs (\ding{176}).

\begin{figure}
    \centering
    \includegraphics[width=0.98\linewidth]{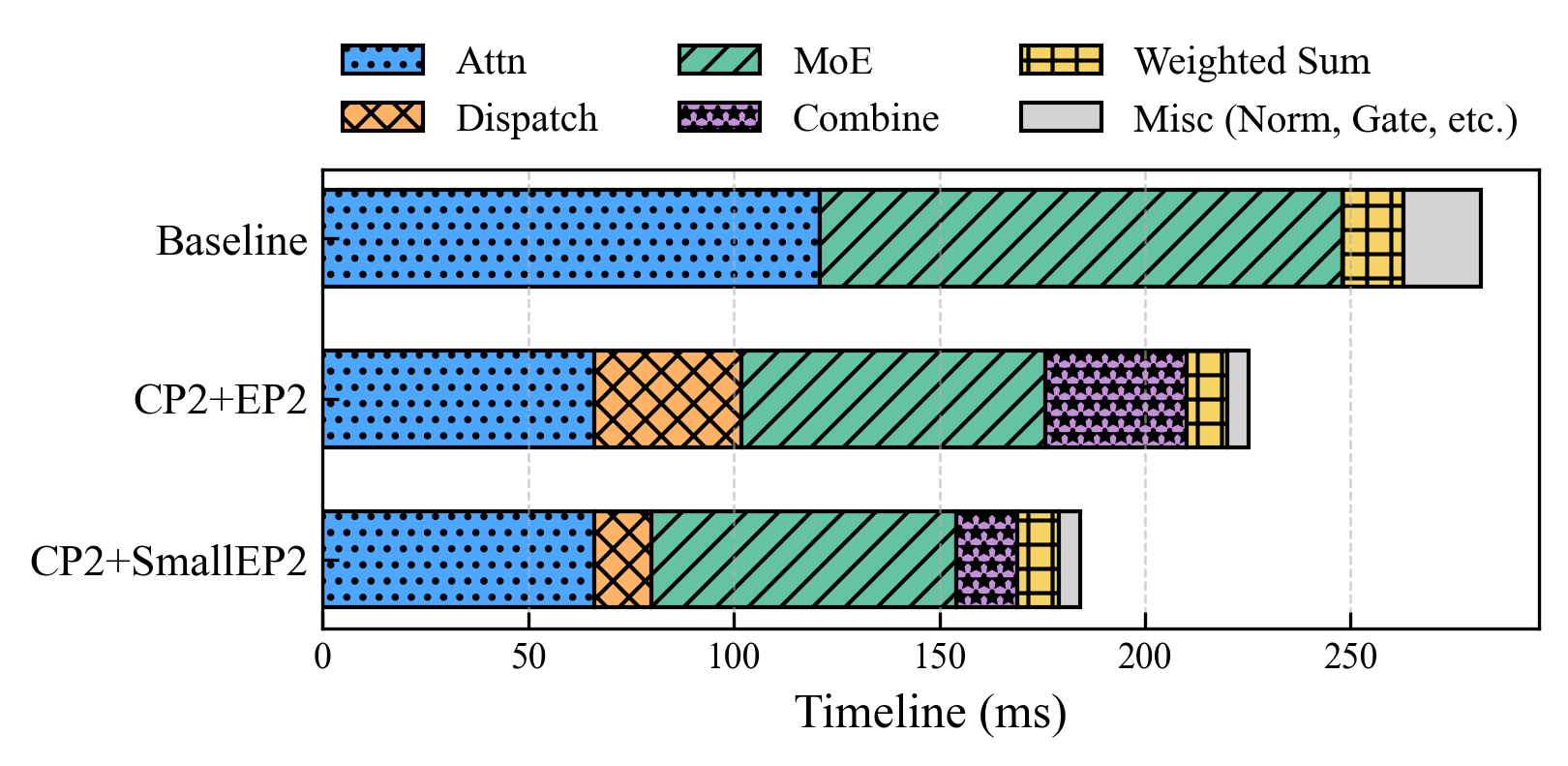}
    \caption{Top-down comparison of parallelism designs on a 20K-token prefill for a single decoder layer (one Attention + one MoE). Baseline: single-GPU sequential. CP2+EP2: StripedAttention (CP=2) with standard expert parallelism (EP=2). CP2+SmallEP2: StripedAttention with our optimized expert parallelism that reduces inter-GPU communication by roughly 50\%.}
    \label{fig:topdown}
\end{figure}

Implementing StripedAttention with standard expert parallelism confirms that distributed prefill can effectively reduce computation time. As shown by the top-down microbenchmark of a 20K-token prefill on a single decoder layer in Figure~\ref{fig:topdown}, standard CP2+EP2 reduces end-to-end layer latency by 21\% over the single-GPU baseline. However, EP still incurs substantial PCIe traffic: Dispatch and Combine account for roughly 31\% of the layer time in local deployments without GPU--GPU P2P support. To mitigate this bottleneck, in Section~\ref{sec:prefill:parallel} we propose SmallEP, which redesigns the expert-parallel exchange and reduces transferred MoE communication volume by approximately 50\%. This further lowers end-to-end layer latency by 18\% over standard EP in the same 20K-token prefill microbenchmark. With SmallEP, the 2-GPU distributed stream-loading prefill (DSLP) achieves a 1.64$\times$ throughput gain over single-GPU SLP.

\vspace{-0.7em}
\subsubsection{Concurrency} 
Local deployments face markedly different concurrency profiles  from what is evaluated and optimized for on cloud platforms\cite{li2023alpaserve,kwon2023efficient,zheng2024sglang,shoeybi2019megatron}. Given inherently low parallelism, two common strategies help—but with trade-offs: (1) Chunked prefill\cite{agrawal2023sarathi} partitions long prompts so prefill and decode can interleave, which reduces mutual blocking between a lengthy prefill and concurrent decodes on hybrid CPU–GPU systems. However, segmentation induces frequent kernel transitions and short compute windows, destabilizing per-token latency (TPOT) for decode and often increasing time-to-first-token (TTFT) for prefill versus isolated execution due to altered scheduling. (2) Batched decode and continuous batching\cite{yu2022orca} coalesce decode steps across requests to execute jointly. While highly effective in the cloud—large batches saturate expert-activation paths and raise compute utilization—the gains diminish locally. With small batch sizes and sparse expert activation, batching multiplies the number of active experts per step, amplifying memory-bandwidth pressure; consequently, each request experiences worse TPOT compared to non-batched local execution. In Section~\ref{sec:concurrency} we provide intra-node prefill-decode disaggregation and dual-batch attention-MoE overlap to overcome these two types of contentions.

\vspace{-0.7em}
\subsubsection{FP8 Block Scaling Format}

\begin{figure}[t]
    \centering

    \begin{subfigure}[t]{0.48\linewidth}
        \centering
        \includegraphics[width=\linewidth]{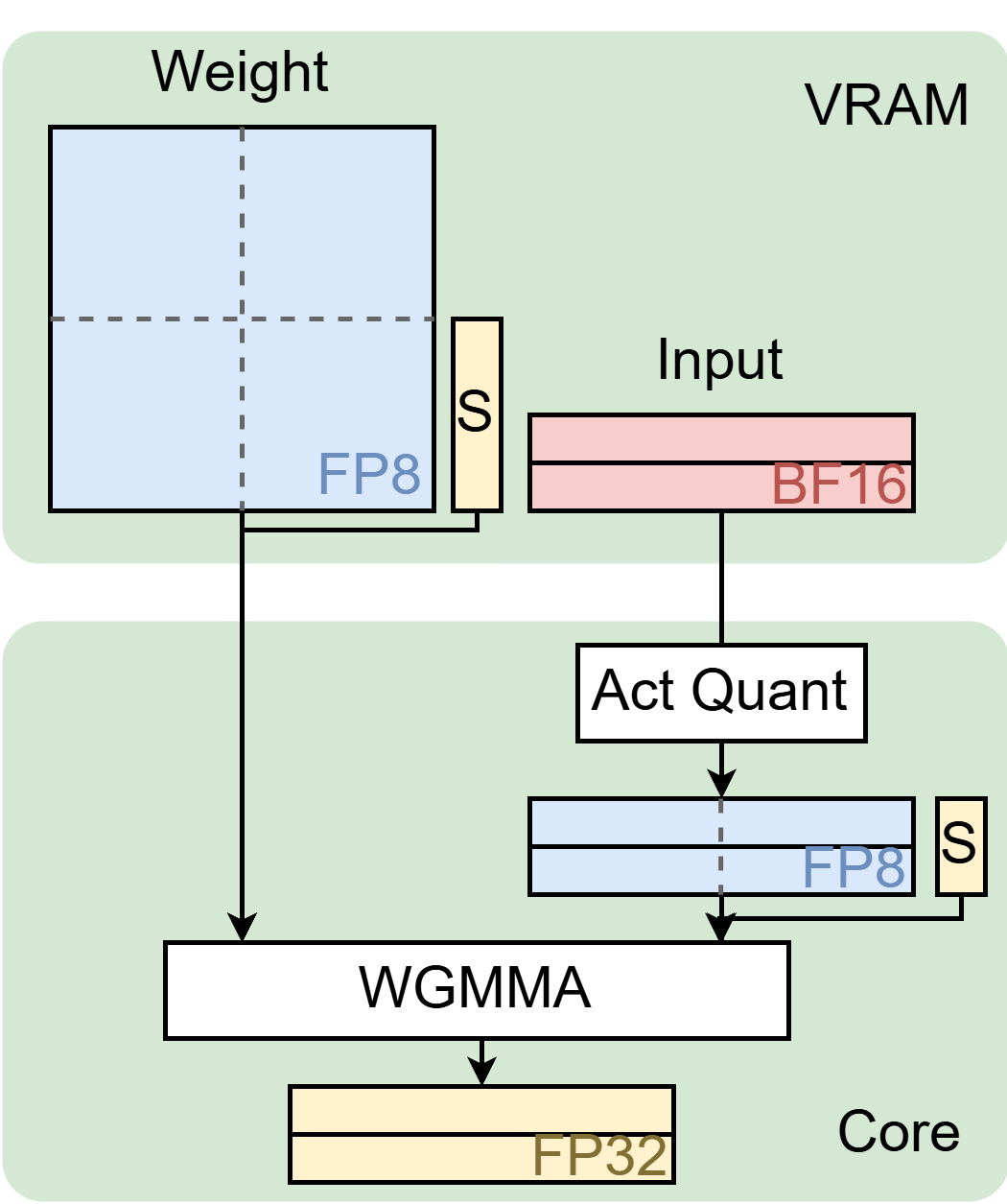}
        \caption{Example FP8 GEMM execution path on a modern GPU. 
        }
        \label{fig:fp8:gpu}
    \end{subfigure}
    \hfill
    \begin{subfigure}[t]{0.48\linewidth}
        \centering
        \includegraphics[width=\linewidth]{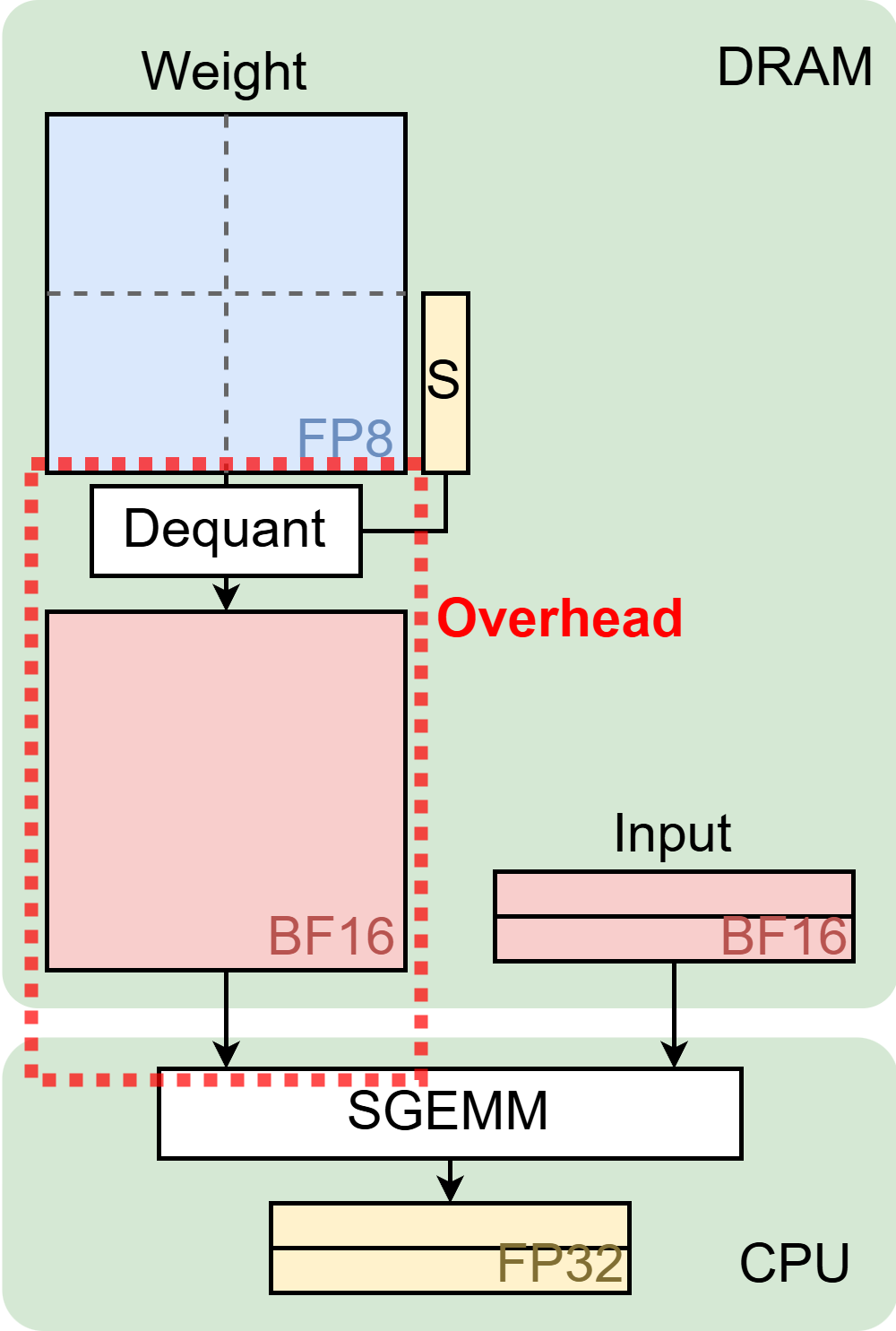}
        \caption{Example FP8 GEMM execution path on a CPU. 
        }
        \label{fig:fp8:cpu}
    \end{subfigure}

    \caption{FP8 GEMM execution on GPU versus CPU. “S” denotes per-tensor or per-block scale factors. On CPU, FP8 weights and scales are dequantized to BF16 prior to GEMM; the dequantization and doubled memory footprint are highlighted by the red box.}
    \label{fig:fp8}
\end{figure}

Modern models increasingly train and serve with native low-bit formats such as FP8 and FP4\cite{nvfp8, mxfp8, wang2025optimizing}, expanding parameter capacity while reducing memory pressure and preserving capability and generality\cite{kimiteam2025kimik2openagentic, guo2025deepseek, liu2024deepseek, peng2023fp8, micikevicius2022fp8, fishman2024scaling}. FP8 typically uses per-tensor or per-block scaling to map BF16 ranges into E4M3 (4-bit exponent, 3-bit mantissa), restoring dynamic range via shared FP32 scales. On modern GPUs, hardware support\cite{NVIDIA_Hopper_Whitepaper_2022} and optimized kernels\cite{deepseek2025deepgemm} enable a straightforward FP8 GEMM path (Figure~\ref{fig:fp8:gpu}): FP8 weights and scales are loaded from VRAM, activations are quantized to FP8, and WGMMA executes native FP8 GEMM with FP32 accumulation. 

In contrast, CPUs lack native FP8 units; hybrid systems run quantized\cite{r1gguf, v3gguf} or upcasted BF16 models\cite{r1bf16, v3bf16} to execute on CPUs\cite{chen2025ktransformers}. Quantization can degrade precision and stability, while upcasting doubles memory requirements. As shown in Figure~\ref{fig:fp8:cpu}, FP8 weights must be dequantized to BF16 before SGEMM, inflating DRAM capacity and memory traffic—the red boxed region marks this dequantization overhead.

To overcome these limitations and enable original-precision FP8 inference on CPUs, we develop an AVX-512–optimized FP8 GEMV kernel tailored for online FP8 MoE serving, As detailed in Section~\ref{sec:fp8kernel}. Our FP8 kernel delivers 4–5× lower latency and higher effective memory bandwidth than existing BLAS libraries, making native CPU FP8 inference practical without quantization/dequantization.

\section{System Design} 

\paragraph{Hardware and model target.} Our goal is to meet cloud-level SLOs on a single, cost-efficient local node. The system prioritizes price/performance: dual-socket, value-tier server CPUs (e.g., AMD EPYC 9355–class) supplies enough DRAM capacity and bandwidth, while 1–2 consumer GPUs (e.g., RTX 5090–class) provide the bulk of compute at low cost per TFLOP. On the model side, we primarily target intact, original-precision large MoE models whose full weights fit in host DRAM, especially DeepSeek-V3-class models and similar state-of-the-art public MoE models.
\subsection{Stream-loading Prefill}
\label{sec:prefill}

\begin{figure}
    \centering
    \includegraphics[width=0.95\linewidth]{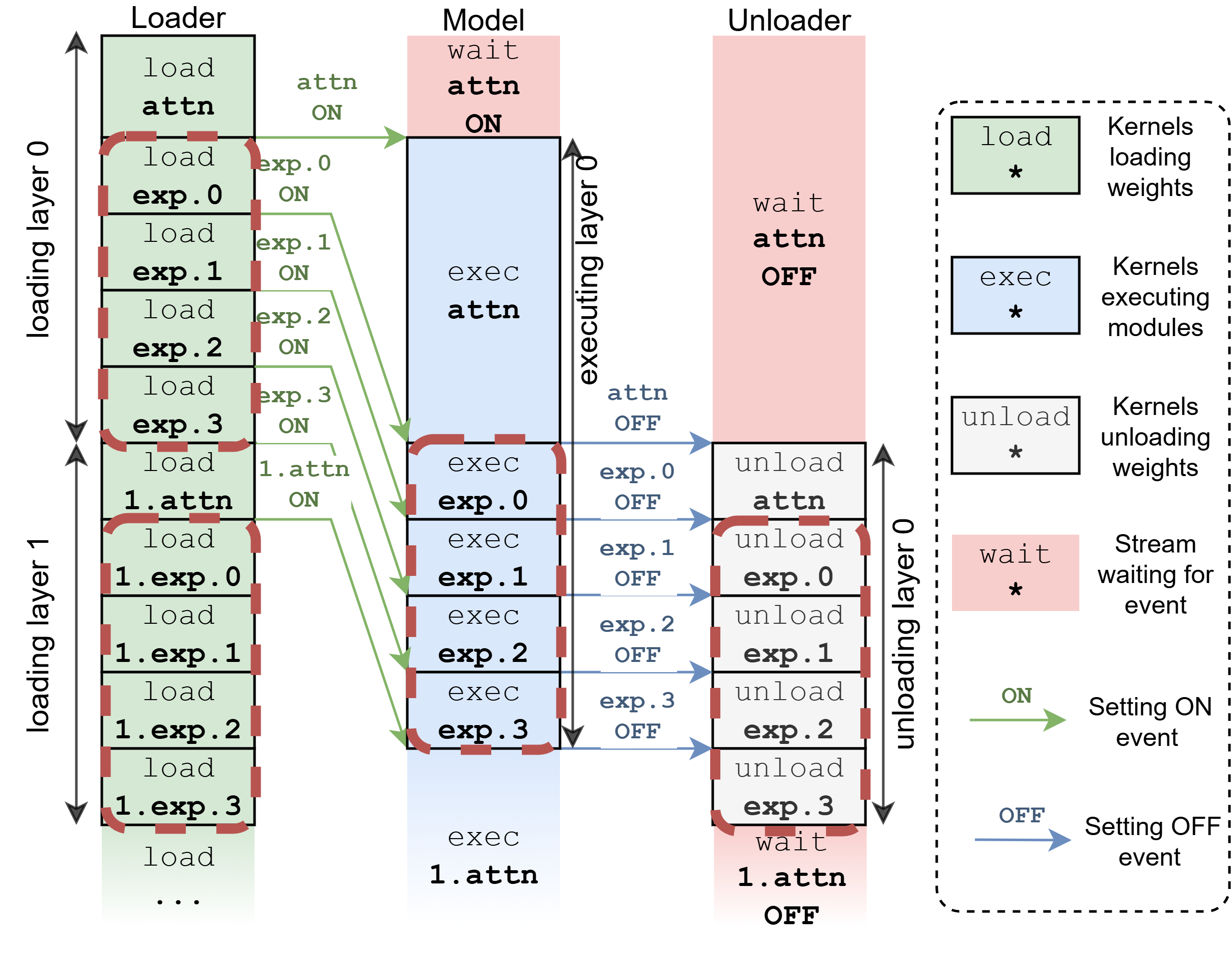}
    \caption{Stream-loading prefill pipeline with three concurrent threads. The loader (left, green) loads weights for modules (attn \& exp.i) and sets ON events. The model thread (center, blue) wait ON events, executes modules and set OFF events. The unloader (right, gray) wait OFF events then unloads weights to free GPU memory. Arrows indicate event signaling and dependency ordering. Dark red boxes indicates weights that are managed by experts ring buffer.}
    \label{fig:slp}
\end{figure}

To maximize GPU utilization and reduce TTFT, we prefill exclusively on GPUs with a computation–communication overlap design inspired by weight-offloading systems. Unlike prior approaches that rely on modified routing or compression, our design targets overlapping operations and maximizing prefill throughput for intact large MoE models. 

\vspace{-0.7em}
\paragraph{Stream-loading prefill scheme.} 
We implement prefill as a three-thread (as well as three CUDA streams) pipeline—loader, model, and unloader—coordinated at a sub-layer granularity, with event tuples that each pairs a CUDA event (device-side) with a host threading event (CPU-side). The loader runs concurrently with the model thread, proactively staging module weights from DRAM into GPU global memory. For each module, the loader signals readiness by setting its ON events, which the model thread waits on before executing that module. After a module finishes, the model thread sets the corresponding OFF events to indicate its weights can be released. The unloader waits on OFF events and, once signaled, evicts the associated weights from GPU memory. This design overlaps weight I/O with computation and minimizes stalls in computation, and bounds VRAM use by actively reclaiming weights after their execution. 

\vspace{-0.7em}
\paragraph{Mitigating frequent memory operations.} A key engineering challenge is the sheer number of weight tensors. In DeepSeek-V3, 58 MoE layers × 256 experts × 3 FP8 linear modules yields 44.5K tensors; naïvely allocating and freeing them incurs substantial cudaMalloc/cudaFree overhead. We therefore use a manually managed experts ring buffer shared across layers, as marked in dark red boxes in Figure~\ref{fig:slp}. Its length adapts to the compute–communication balance and the peak memory footprint of long-context prefill. For short prompts (transfer-dominated), we set the length to the number of experts of a layer to maximize preload lookahead and saturate host–device bandwidth. Beyond a prompt-length turning point (e.g. 50K), computation becomes the bottleneck and intermediate tensors dominate VRAM; we shrink the length to 2 (ping–pong) to maintain overlap while capping memory use.

\vspace{-0.7em}
\paragraph{Loader and unloader strategy.} The loader must decide whether the next module’s weights fit in VRAM. For tensors not covered by the ring buffer, we probe available device memory before loading. These tensors are relatively small; a fixed budget of about 1~GB suffices to stage one layer’s dense weights. The loader streams weights until that budget would be exceeded, then blocks until the unloader frees space.
The unloader follows different policies for different tensor classes. For expert weights managed by the ring buffer, the authoritative copy always remains in host memory, so unloader simply frees or reuses the corresponding GPU-side ring-buffer slot rather than copying the weights back to the CPU. For other tensors, such as the QKV projection weights, which are few in number and irregular in shape and are therefore not managed by the experts ring buffer, the unloader simply releases their temporary GPU allocations after use.

Compared with more implicit memory-management and prefetching mechanisms such as CUDA Unified Virtual Memory and \texttt{cudaMemPrefetchAsync}, our design uses explicitly orchestrated loader/model/unloader threads to obtain predictable memory usage and precise control over computation-transfer overlap. This choice does introduce additional engineering complexity, but the added logic is mostly confined to reusable ring-buffer management and synchronization primitives, making the design straightforward to adapt across different large MoE models.

\subsection{Distributed Stream-loading Prefill} \label{sec:prefill:parallel}

In few-GPU settings typical of local deployments, we distribute prefill with a combination of context parallelism (CP) and expert parallelism (EP) across GPUs to maximize hardware utilization and shorten time-to-first-token (TTFT); we refer to this distributed version as DSLP. The design targets low parallelism sizes and inferior interconnect regimes by carefully balancing computation and communication, co-optimizing data layout to reduce inter-GPU transfers. 

For attention, we adopt the zig-zag variant of striped attention \cite{brandon2023striped, zhuzilin_ring_flash_attention}, which achieves load balance across GPUs while simplifying implementation.

\begin{figure}
    \centering
    \begin{subfigure}[t]{0.45\linewidth}
        \centering
        \includegraphics[width=\linewidth]{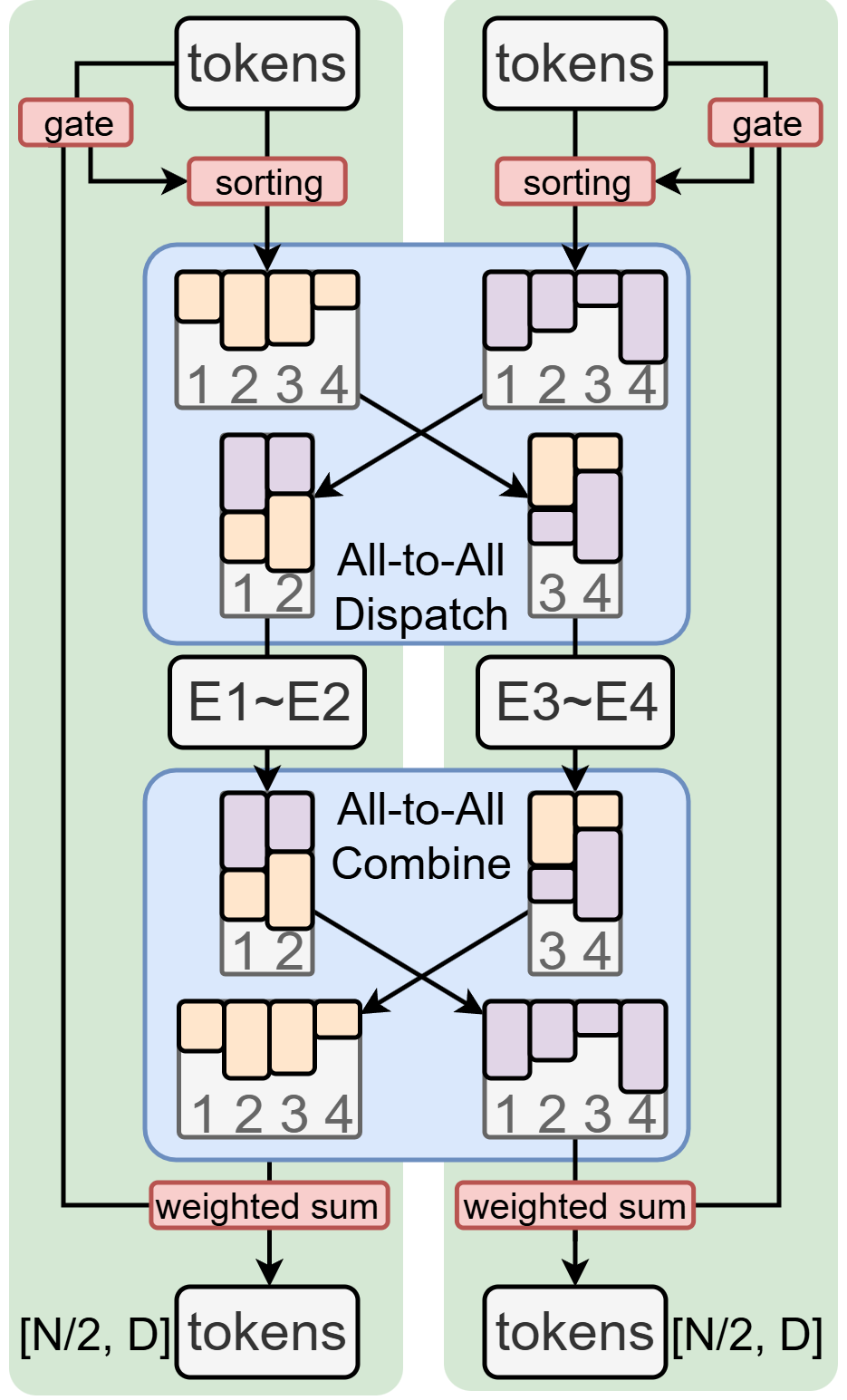}
        \caption{Standard EP workflow.}
        \label{fig:ep:ep}
    \end{subfigure}
    \hfill
    \begin{subfigure}[t]{0.46\linewidth}
        \centering
        \includegraphics[width=\linewidth]{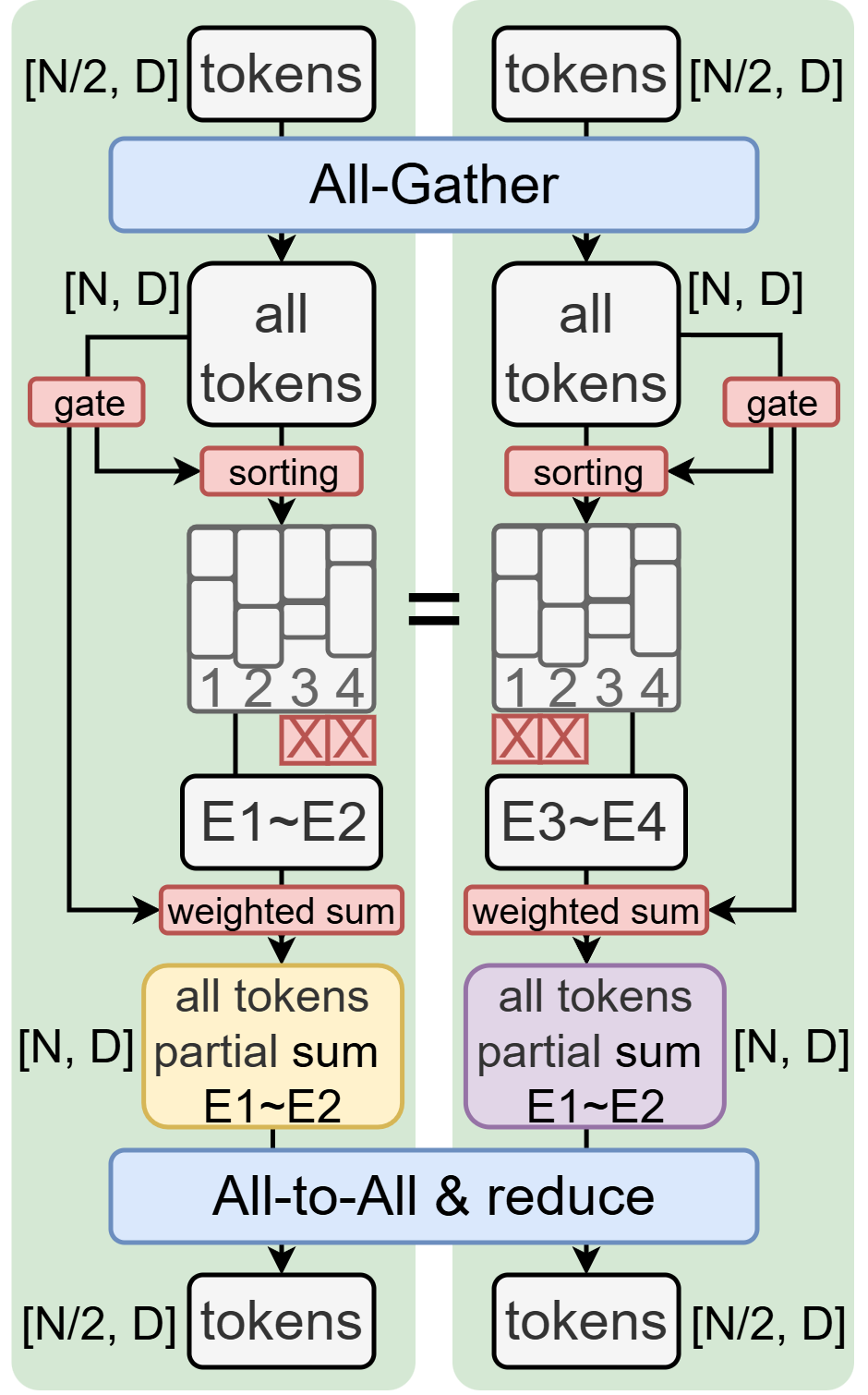}
        \caption{SmallEP workflow.}
        \label{fig:ep:smallep}
    \end{subfigure}
    \caption{Workflow comparison between standard expert parallel and SmallEP optimization under an EP size=2 configuration. N is the number of tokens on all devices, D is hidden size.}
    \label{fig:lowep}
\end{figure}

We optimize EP for small EP sizes to reduce communication overhead on our few-GPU scenario with commodity interconnects. In the standard EP workflow (Figure~\ref{fig:ep:ep}), tokens are routed by the gate, sorted by expert, and then exchanged via an All-to-All dispatch so each device gathers the tokens for its local expert group. The expected volume communicated per device pair is $\cfrac{N\cdot A\cdot D}{S}$, where $N$ is the total number of tokens across devices, $A$ is the number of activated experts per token, $S$ is the EP size, and $D$ is the hidden-state size; the combine phase mirrors this cost. When $S$ is large, traffic is evenly spread across links and latency can be effectively overlapped. However, with small $S$—typical in local deployments without high-bandwidth interconnects like NVLink—the per-link volume becomes concentrated and the transmission cost dominates, as depicted in Figure~\ref{fig:topdown}. 

\vspace{-0.7em}
\paragraph{SmallEP design.} To reduce communication in small EP regimes, we first All-Gather the unsorted tokens so that every rank holds a full copy of all tokens, then perform gate evaluation and token sorting locally on this replicated batch. Each rank retains tokens assigned to its local experts and discards the rest before expert computation. After experts finish, we apply a per-rank partial weighted sum to produce a reduced tensor of shape $[N, D]$, perform an All-to-All on these partial results, and then complete the local reduction to obtain the final output. 

This scheme fixes the All-Gather traffic at $N\cdot D$ bytes per link and limits the All-to-All to the partially reduced $N\cdot D$-sized tensors. Compared to standard EP's communication volume $\cfrac{N\cdot A\cdot D}{S}$, we deduce that when the EP size $S$ is no larger than the number of activated experts $A$, SmallEP reduces peak per-link bandwidth and improves end-to-end communication efficiency. 

This tradeoff is favorable in the small-EP regime we target, especially for long-context prefills over PCIe-class interconnects, where dispatch/combine communication becomes a dominant cost. SmallEP does incur redundant local gating and sorting because each rank processes a full copy of all tokens before discarding non-local expert assignments. However, our profiling shows that this overhead---included in the ``misc'' component of Figure~\ref{fig:topdown}---remains below 10\,ms per layer (\(<5\%\) of end-to-end latency), and is therefore small relative to the communication overhead that SmallEP eliminates.

Expert load imbalance is a well-known concern in large-scale EP deployments. Howerver, in our target small-EP regime, each device serves a relatively large set of experts rather than only a few, so routing skew is averaged out more effectively within each device. This makes stragglers from expert imbalance much less severe than in traditional large-EP settings.

\subsection{Concurrency-Targeted Optimizations}
\label{sec:concurrency}

\begin{figure}
    \centering
    \includegraphics[width=0.65\linewidth]{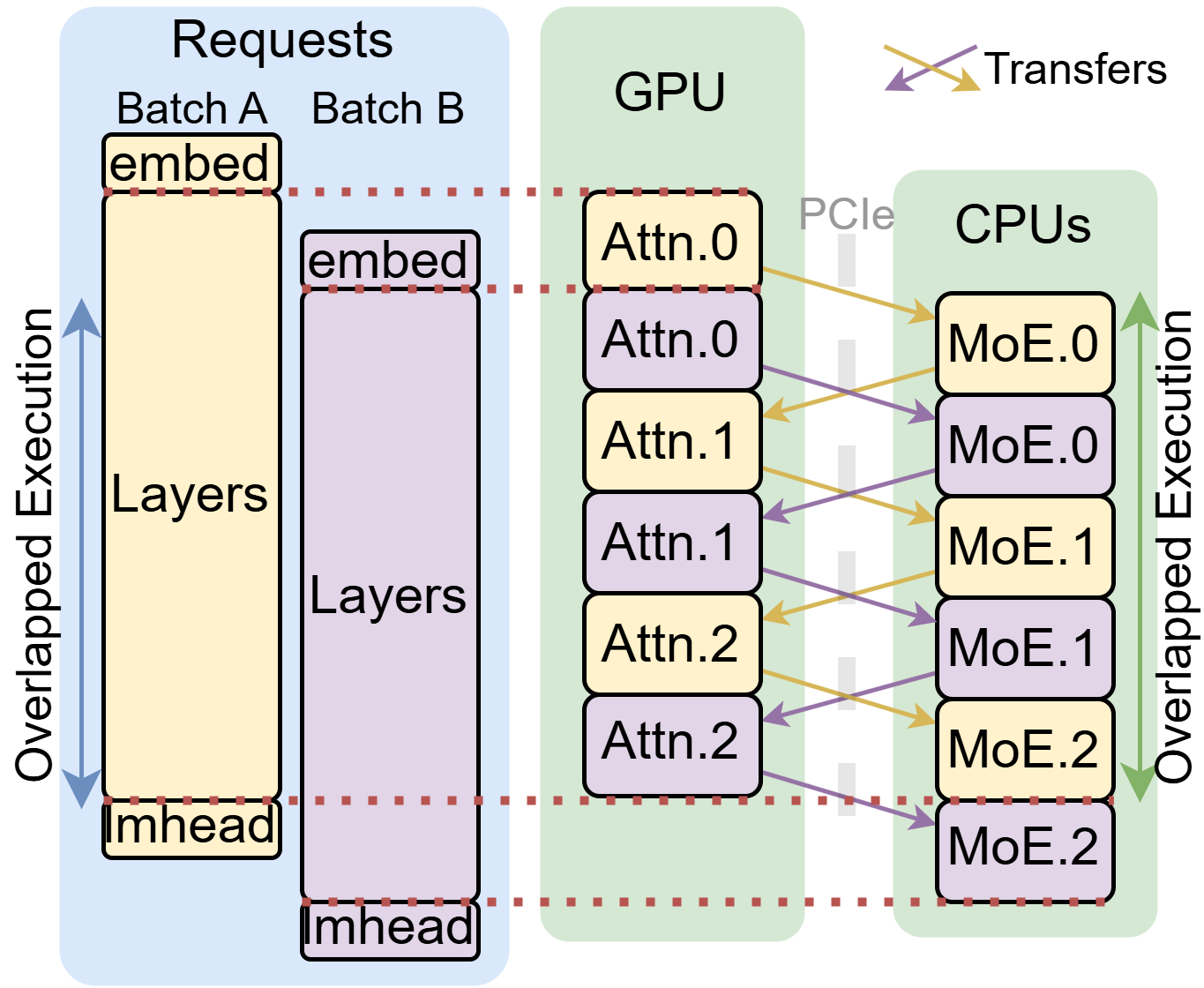}
    \caption{Dual-batch attention-MoE parallelism scheme for simultaneous inference of two batches A and B. "embed" stands for input token embedding linear layer; "lmhead" stands for output language model head linear layer.}
    \label{fig:2seq}
\end{figure}

\subsubsection{Dual-Batch Attention-MoE Overlap}

Batched decoding presents a fundamental bottleneck on CPU-centric systems due to independent MoE activation patterns. As the number of concurrent requests increases, the set of activated experts expands multiplicatively, creating proportional increases in memory bandwidth demand that quickly saturate available resources.

On the other hand, hybrid CPU-GPU inference reveals a sequential execution pattern where CPU and GPU components alternate but cannot overlap due to compute dependencies. When the CPU processes MoE layers, the GPU remains idle, and vice versa. Our measurements show that attention processing requires approximately 350µs while MoE computation takes 450µs per layer, creating significant idle time that represents wasted computational capacity.

To address both the concurrency limitations and hardware underutilization, we propose a dual-batch interleaving execution scheme. As illustrated in Figure~\ref{fig:2seq}, we create two threads as well as two CUDA streams (A and B, execution flow represented by arrows) for concurrent request processing. The execution pattern interleaves two streams of attention (on the GPU) and MoE computation (on the CPU), creating a tightly coupled pipeline that minimizes idle time. The only synchronization bottlenecks occur during GPU-exclusive operations such as token embedding and model head computation, but these contribute minimal overhead (<5ms) compared to the 61 decoder layers that dominate execution time (approximately 50ms per token).

\subsubsection{Intra-node Prefill-Decode Disaggregation}

Building on stream-loaded prefill, we push prefill compute fully onto the GPU with minimal CPU time consumed. This enables disaggregation within a single node: we dedicate one GPU to prefill while the other handles decoding or chunked prefill, thereby isolating prefill tasks from decode ones, increasing concurrency and reducing contention impacts.

To avoid duplicated storage of weights in DRAM, we share a single resident copy of model weights in our system with the SLP or DSLP pipelines. Inter-process communication uses the experts ring buffer to achieve zero memory copies and minimizes DRAM bandwidth consumption. During disaggregated SLP+decode, Although weights still necessarily traverse the CPU–GPU path over a PCIe 5.0 x16 link, this cost is modest relative to the >1 TB/s DRAM bandwidth and does minimal impact on decode performance.

Intra-node prefill–decode disaggregation, combined with DSLP, expands the scheduling design space beyond plain chunked prefill. Our policy is: use chunked prefill for short requests (e.g., <2K tokens). For long prefill requests, if there are no decode requests within a short window (e.g., 5 minutes), route to DSLP to maximize throughput; otherwise, route to disaggregated SLP to preserve decode latency. Prefill requests are served first-come, first-served, while decode requests are dynamically batched up to a target size (e.g., Y=6) to balance latency and efficiency.

\subsection{AVX-512 FP8 GEMV Kernel}
\label{sec:fp8kernel}

At the core of our system lies an FP8 matrix-vector (GEMV) kernel optimized for modern AVX-512 CPUs where modern GPU-style FP8 units are not available. The design targets MoE inference workloads, where extreme throughput and memory efficiency are critical.

\vspace{-0.7em}
\subsubsection{FP8 Format and Dequantization Strategy}

DeepSeek-V3 adopt the E4M3FN format—1 sign bit, 4-bit exponent, and 3-bit mantissa without NaNs—as the FP8 variant for weight storage\cite{liu2024deepseek}. Unlike GPUs, CPUs lack native FP8 processing units\cite{intelia32}. Therefore, we implement E4M3FN dequantization via integer bit manipulations. 

\begin{figure}
    \centering
    \includegraphics[width=0.95\linewidth]{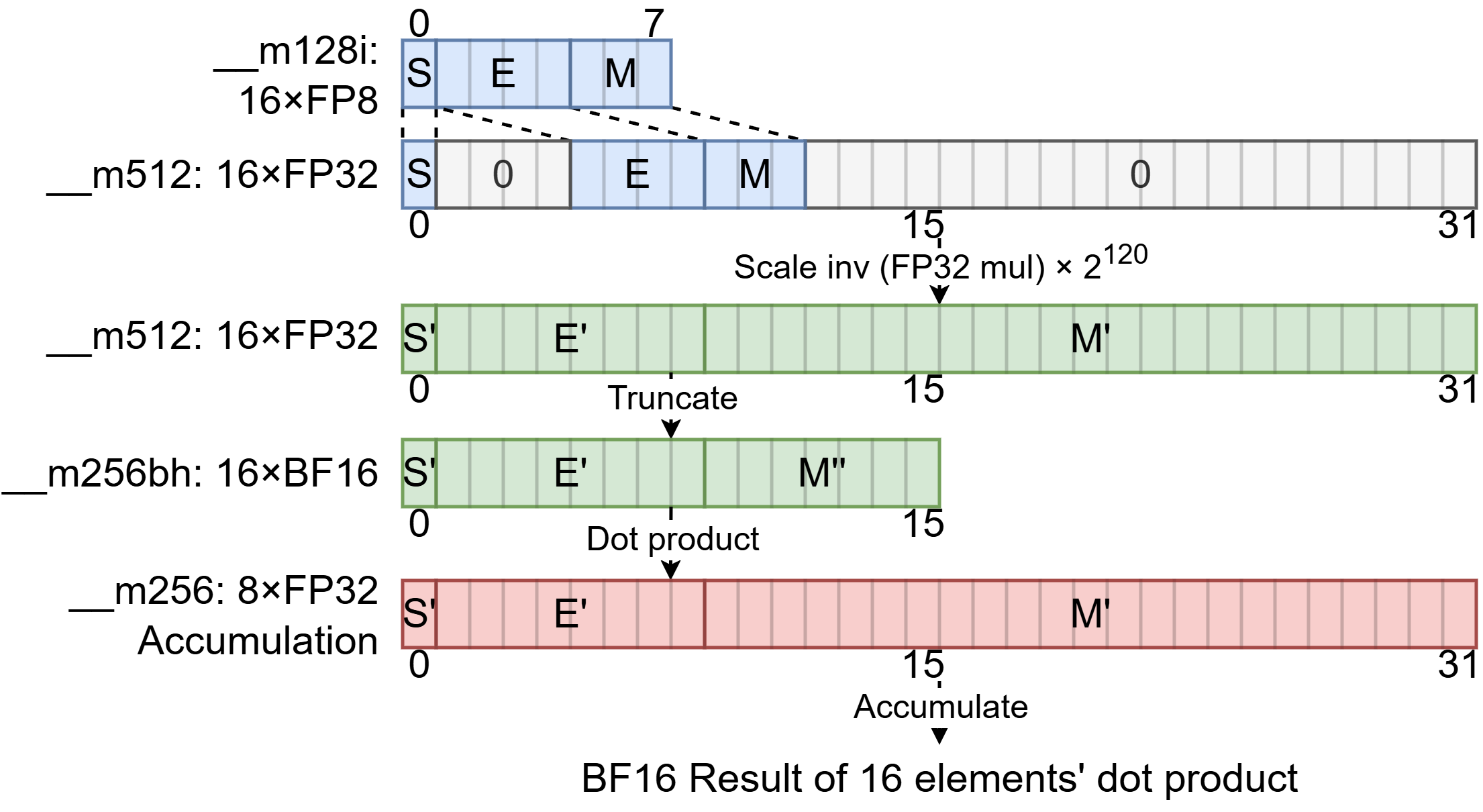}
    \caption{Baseline FP8-BF16 dot-product kernel in AVX512.}
    \label{fig:fp8-cons}
\end{figure}

We implemented a standard FP8-BF16 dot-product kernel, as illustrated in Figure~\ref{fig:fp8-cons}, it requires:
\begin{enumerate}
    \item Expanding FP8 to FP32 (without fixing exponent bias diff),
    \item Applying FP32 scale inverse and exponent bias fix,
    \item Truncating to BF16,
    \item Performing BF16 dot-product.
\end{enumerate}

This approach is inspired by the one that is adopted by Chitu engine\cite{chitu}. Although precise, it involves over 15 vector instructions and suffers from pipeline pressure due to FP32 register density (16 FP32 elements per 512-bit register). On AVX-512, this limits vectorized throughput and results in an estimated latency of $\sim$36 cycles, and throughput of 12 cycles per iteration (per 16 elements).
\begin{figure}
    \centering
    \includegraphics[width=0.95\linewidth]{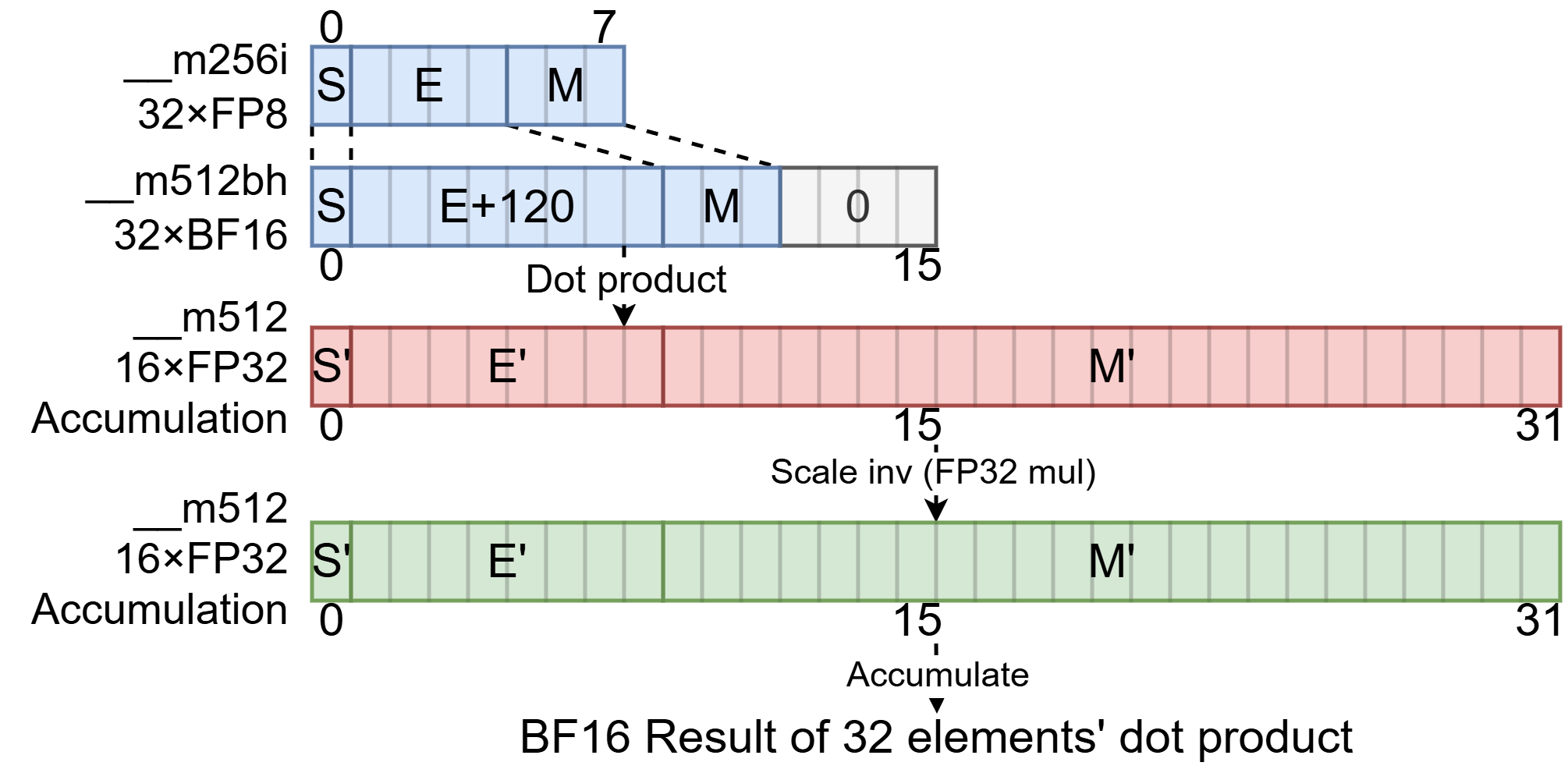}
    \caption{Our post-scaling kernel of FP8-BF16 dot product.}
    \label{fig:fp8-later}
\end{figure}

Our optimized post-scaling design avoids FP32 in the hot path. As illustrated in Figure~\ref{fig:fp8-later}, we:
\begin{enumerate}
    \item Expand FP8 weights directly to BF16,
    \item Load BF16 activations into \texttt{\_\_m512bh},
    \item Execute \texttt{vdpbf16ps} for fused BF16-BF16 dot products into FP32 accumulators,
    \item Apply post-scaling to the accumulated FP32 result.
\end{enumerate}

This optimized execution path circumvents the register pressure caused by frequent FP32 use. By working directly in BF16 for both weights and activations, we effectively double AVX-512 vector slot utilization—from 16 FP32 to 32 BF16 elements per register—maximizing instruction throughput and widening the execution pipeline. An estimated latency of 27 cycles and throughput of 4.96 cycles per iteration (per 32 elements) are deduced.

The numerical impact of this transformation is minimal: the end-to-end GEMV operation exhibits only 0.0017 as the 95 percentile of the L1 differences of all elements when compared to PyTorch’s reference BF16 GEMV implementation on GPUs. This suggests the BF16 intermediate format is sufficient to preserve inference fidelity for most workloads. Beyond this kernel-level result, Section~\ref{sec:eval:quality} further evaluates end-to-end model quality.

Measured on a our dual-socket AVX-512 CPU platform, the post-scaling kernel achieves 815~GB/s memory bandwidth. Supporting CPU FP8 inference even in the absence of dedicated FP8 hardware.

\vspace{-0.7em}
\subsubsection{Tiling and Scale-Block Optimizations}

\begin{figure}[t]
    \centering
    \includegraphics[width=0.85\linewidth]{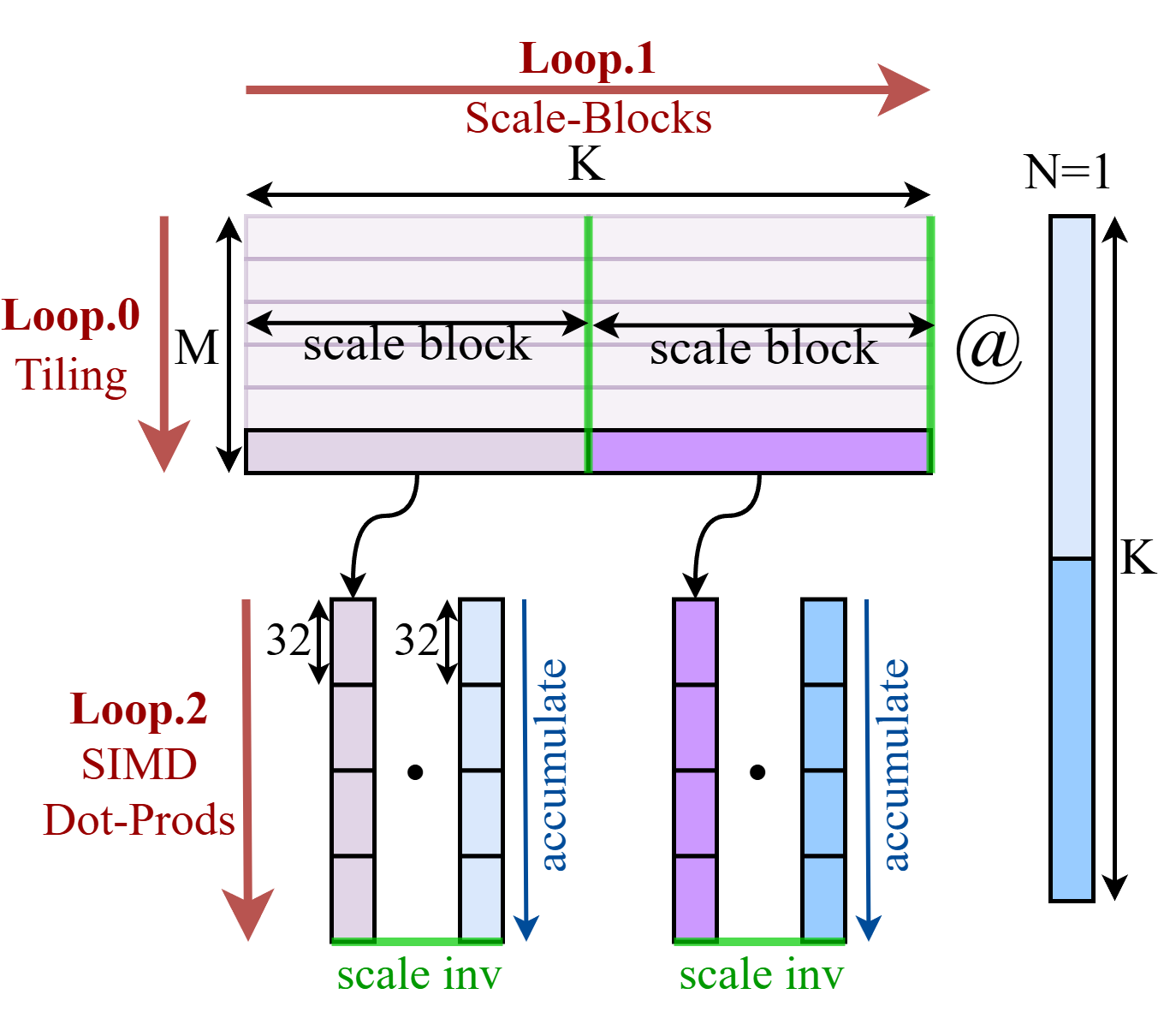}
    \caption{Hierarchical structure of the FP8 GEMV kernel. Computation is organized into \textbf{Tiling} along the M-axis, \textbf{Scale-Block} iteration along the K-axis (128 elements per block), and \textbf{SIMD Dot-Prods} that pipeline accumulations within each block. A single scale factor is loaded per block and applied after all accumulations, reducing scaling overhead and sustaining high bandwidth.}
    \label{fig:tiling}
\end{figure}

Building upon the post-scaling optimization that lifted bandwidth to 815 GB/s, we further refine the GEMV kernel by restructuring the loop hierarchy. Specifically, we decompose the computation into three nested levels (see Figure~\ref{fig:tiling}): (1)~\emph{Tiling} (M-axis): the matrix is partitioned into tiles that do not cross scale-block boundaries, ensuring clean alignment between data layout and scaling granularity. (2)~\emph{Scale-Blocks} (K-axis): the K dimension is iterated in fixed-size blocks (typically 128 elements), each governed by a single FP8 scale factor. Instead of applying the inverse scale per sub-iteration, we accumulate all partial dot-products first and apply the scale\_inv once per block—reducing the number of multiplications by 4× compared to the baseline post-scale kernel. (3)~\emph{SIMD Dot-Prods}: within each block, unrolled inner loops perform fused multiply-accumulate operations in SIMD registers, pipelining partial sums before applying the shared scale factor.

This hierarchy minimizes redundant scaling operations while retaining numerical accuracy, maximizing bandwidth utilization. As a result, the optimized kernel sustains up to 947 GB/s memory bandwidth—surpassing both the baseline and prior post-scaling design.

\subsection{Fine-grained CPU Parallelism for MoE Inference}
\label{sec:cpu-backend}

\subsubsection{Fine-Grained Synchronization Design}

\begin{figure}[t]
    \centering
    \includegraphics[width=0.7\linewidth]{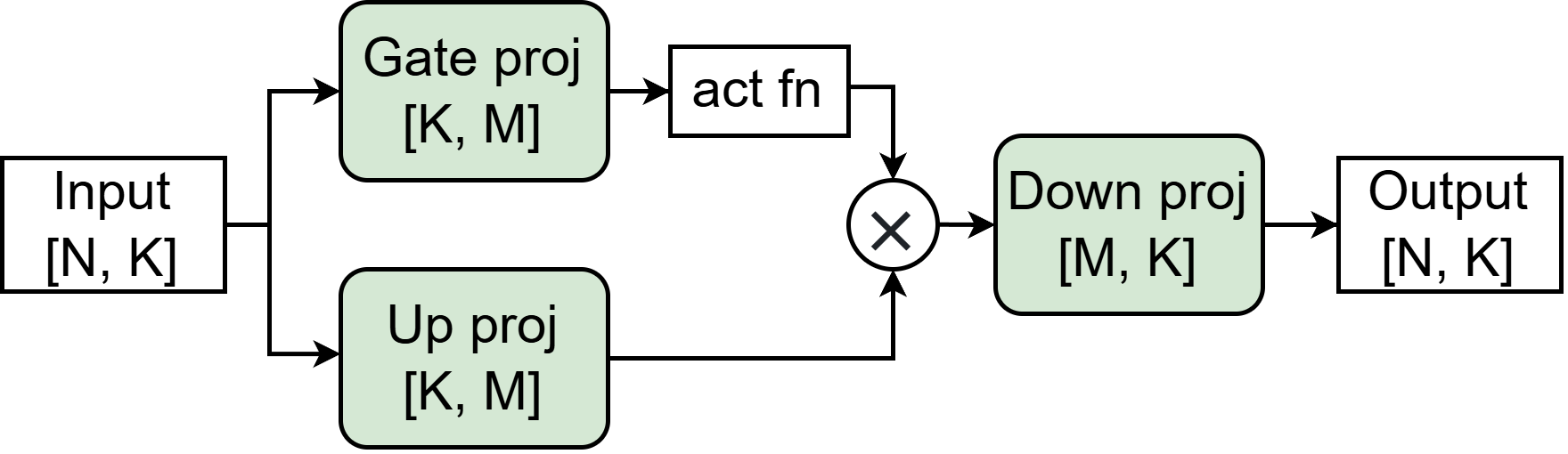}
    \caption{MoE expert compute graph: each expert has gate, up, and down projections. Gate and up can run concurrently; their outputs synchronize before down. N is batch size (N=1 for single-request inference).}
    \label{fig:mlp-flow}
\end{figure}

\begin{figure}
    \centering
    \includegraphics[width=0.8\linewidth]{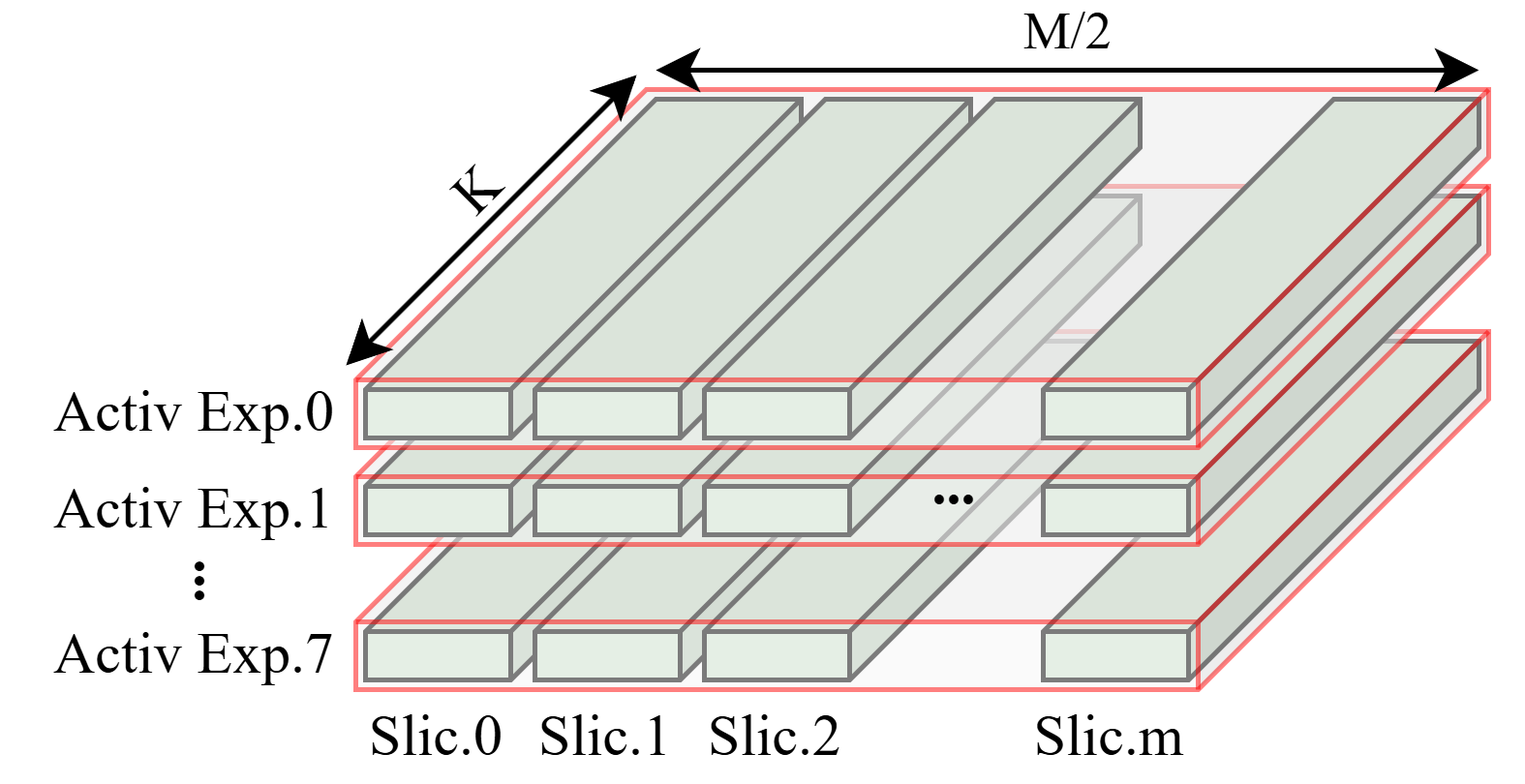}
    \caption{Task partitioning of gate/up on a NUMA node. Output dim M is sliced into m segments per expert, yielding m×8 independent tasks (8 activated experts). Each thread runs a vectorized FP8–BF16 kernel. Red borders mark per-expert barriers that satisfy dependencies for down projection while minimizing synchronization.}
    \label{fig:task-partitioning-gate-up}
\end{figure}

In MoE layers \cite{liu2024deepseek, qwen3}, each expert comprises gate, up, and down projections with the dependencies in Figure~\ref{fig:mlp-flow}. Gate and up are embarrassingly parallel and execute concurrently; their results synchronize before down.

In KTransformers’ CPU backend, stage 1 computes gate and up. As shown in Figure~\ref{fig:task-partitioning-gate-up}, each GEMV is sliced into m segments along the output dimension M, deepening tensor parallelism on the gather axis, while activated experts are evenly distributed across cores within each NUMA node. Because gate and up must be combined (activations and multiplications) before down, naive global barriers could be adopted but are costly. Instead, we use fine-grained per-expert barriers and fuse the combine ops into stage one, cutting synchronization overhead and improving scalability.

\vspace{-0.7em}
\subsubsection{Operator Fusion for Quantized Inference} 

In quantized MoE models, expert weights (gate, up, down) use low-bit formats (e.g., Q4\_K\_M, a mixed blockwise GGUF quantization format with per-block scales/metadata). In the KTransformers configuration we study, \texttt{Q4\_K\_M} is a mixed K-quant storage scheme: \texttt{gate\_proj} and \texttt{up\_proj} use \texttt{Q4\_K}, while \texttt{down\_proj} uses \texttt{Q6\_K} [3]. The expert inference path can be summarized as
\[
\text{input conversion:}\quad \texttt{hidden\_type} \rightarrow \texttt{Q8\_K},
\]
\[
\text{gate/up:}\quad \texttt{Q4\_K}\times \texttt{Q8\_K} \rightarrow \texttt{F32},
\]
\[
\text{activation and elementwise multiply:}\quad \texttt{SiLU}(\cdot)\odot(\cdot)\rightarrow \texttt{F32},
\]
\[
\text{down-input conversion:}\quad \texttt{F32} \rightarrow \texttt{Q8\_K},
\]
\[
\text{down:}\quad \texttt{Q6\_K}\times \texttt{Q8\_K} \rightarrow \texttt{F32},
\]
followed by router-weighted accumulation in \texttt{F32} and conversion back to \texttt{hidden\_type}.

This requires converting activations among FP32, Q8\_0, and BF16 to preserve precision for SiLU and expert aggregation. We fuse these conversions into surrounding kernels: input conversions for gate/up and down are fused into the parallel gate/up projections (stage 1); output conversion is fused into down (stage 2). The result reduces launch overhead and limits synchronization to two global barriers per MoE layer—after gate/up combination and after down.

\section{Evaluations}

Our evaluation is organized compositionally around the targeted deployment setting. The claimed SLOs apply to flagship MoE models on the dual-socket CPU plus 1--2 consumer-GPU class of hardware described in Section~\ref{sec:eval:setup}. We then isolate the major components of the system: Section~\ref{sec:eval:prefill} evaluates SLP/DSLP and SmallEP for long-context prefill; Sections~\ref{sec:eval:decode} and~\ref{sec:eval:decode_q4} evaluate fine-grained CPU parallelism and dual-batch attention--MoE overlap for full-precision and quantized decode; and Section~\ref{sec:eval:concurrency} shows how these components interact under concurrent requests in a single-node deployment. Finally, Section~\ref{sec:eval:fp8kernel} isolates FP8 kernel performance, while Section~\ref{sec:eval:quality} reports end-to-end model quality.

\subsection{Evaluation Platform Setup}
\label{sec:eval:setup}

\paragraph{Hardware Configuration Overview}

Our primary evaluation platform is a hybrid inference node built for cost-effective large-scale model serving. 
Our platform is equipped with two AMD EPYC 9355 processors, 24 DDR5-6400 48GB memory modules distributed across 24 memory channels, yielding a total system memory capacity of 1.15TB and a theoretical aggregate bandwidth of 1228GB/s. Complementing the CPU is two NVIDIA RTX 5090 GPUs with 32~GB of GDDR memory and up to 419 FP8 TFLOPS (with FP32 accumulation). Unless otherwise noted, all experimental results for both our engine and KTransformers are measured on this platform; we also cite KTransformers' published numbers when comparing against their original reported setup. Our empirical evaluation therefore primarily covers this dual-socket, PCIe-attached commodity CPU--GPU node. For lower-end or otherwise different configurations, such as reduced effective DRAM bandwidth or lower-bandwidth PCIe interconnects, we provide discussion in Section~\ref{sec:discussion}.

\vspace{-0.7em}
\paragraph{Model Benchmarks}
Our primary benchmark target is the full-precision FP8 DeepSeek-R1\cite{guo2025deepseek} and Kimi-K2\cite{kimiteam2025kimik2openagentic} model, with the Q4\_K\_M quantized DeepSeek-R1\cite{r1gguf} evaluated alongside them to contextualize the tradeoffs and benefits of our optimizations.

\subsection{Prefill Performance}
\label{sec:eval:prefill}

\begin{figure}
    \centering
    \includegraphics[width=\linewidth]{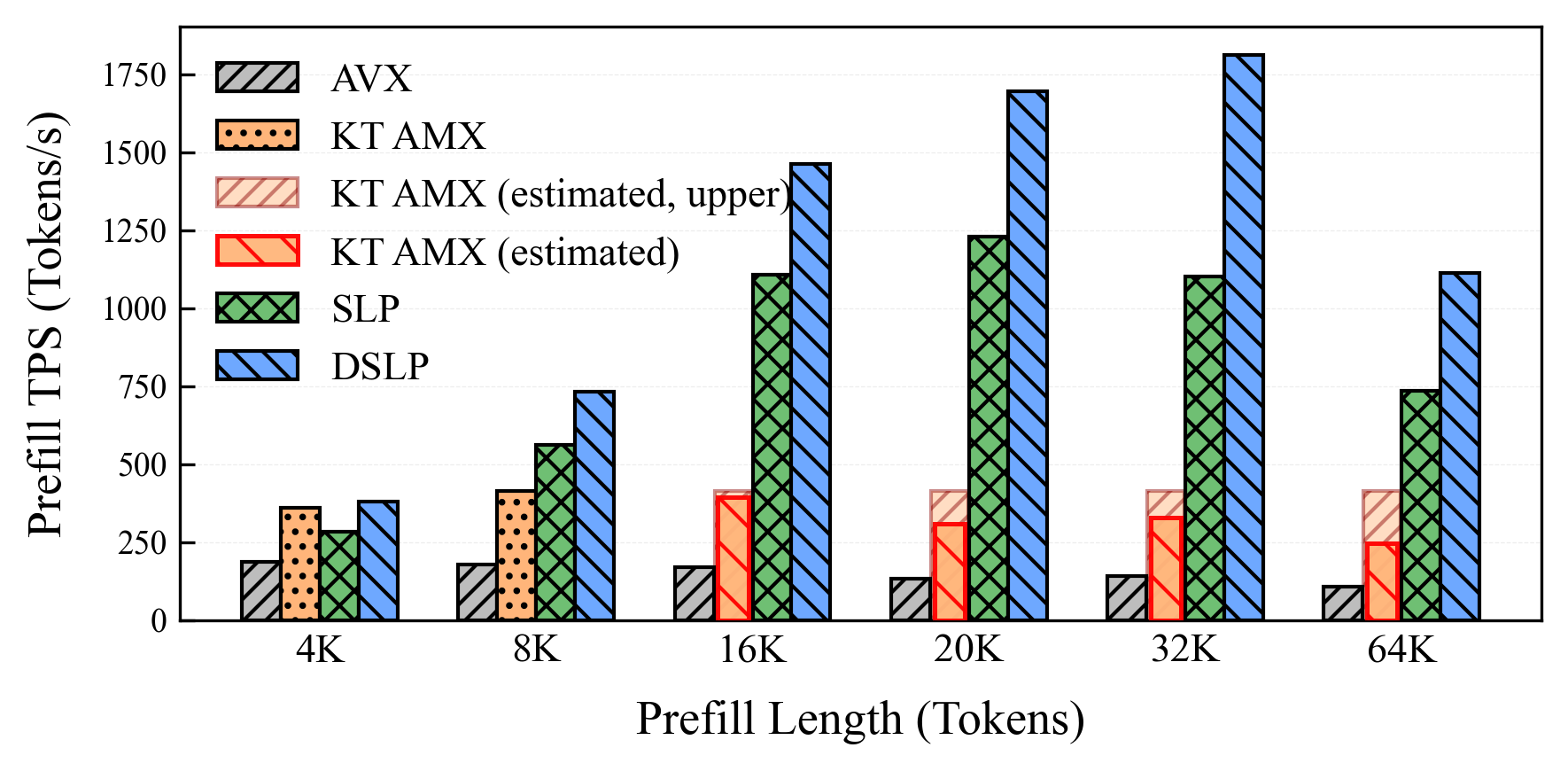}
    \caption{Prefill throughput (tokens/s) for different configurations. We compare CPU AVX baseline (FP8), KTransformers (KT) AMX (INT4), Stream‑Loading Prefill (SLP, FP8), and 2‑GPU Distributed SLP (DSLP, FP8). KT AMX values for prefill lengths >8K are shown as estimates: “estimated” is obtained by linearly scaling the AVX baseline based on the ratio observed at $\leq$8K tokens. “estimated, upper” is an optimistic projection of the measured KT AMX results by assuming both attention time and MoE time scale linearly with prefill length.}
    \label{fig:prefill-tps}
\end{figure}

Figure~\ref{fig:prefill-tps} reports prefill throughput as a function of sequence length for different configurations. On our platform, we evaluate our engine on the original FP8 model with three prefill modes: baseline CPU prefill (AVX), Stream-Loading Prefill (SLP), and its 2-GPU distributed variant (DSLP). For comparison, we also cite KTransformers' published AMX-based prefill results (KT AMX), which were obtained on their own platform rather than ours. Across the long-context regime (8K--64K), SLP and DSLP substantially outperform both AVX and KT AMX. For short contexts (4K), weight transfer dominates SLP and DSLP latency, thus no dominant advantage over KT AMX. As the prefill length increases into the long‑context regime (16–64K), the advantage of SLP becomes dominant. At 20–32K tokens, SLP delivers more than an order‑of‑magnitude speedup over AVX and a 2.8× improvement relative to the estimated KT AMX performance, while DSLP further increases throughput to 1.64× SLP. The “estimated upper” KT AMX bars, obtained by linearly scaling the measured KT AVX throughput, remain far below SLP and DSLP, implying that even an ideal AMX implementation that scales linearly with sequence length cannot match the gains from our SLP and DSLP.

\begin{figure}
    \centering
    \includegraphics[width=\linewidth]{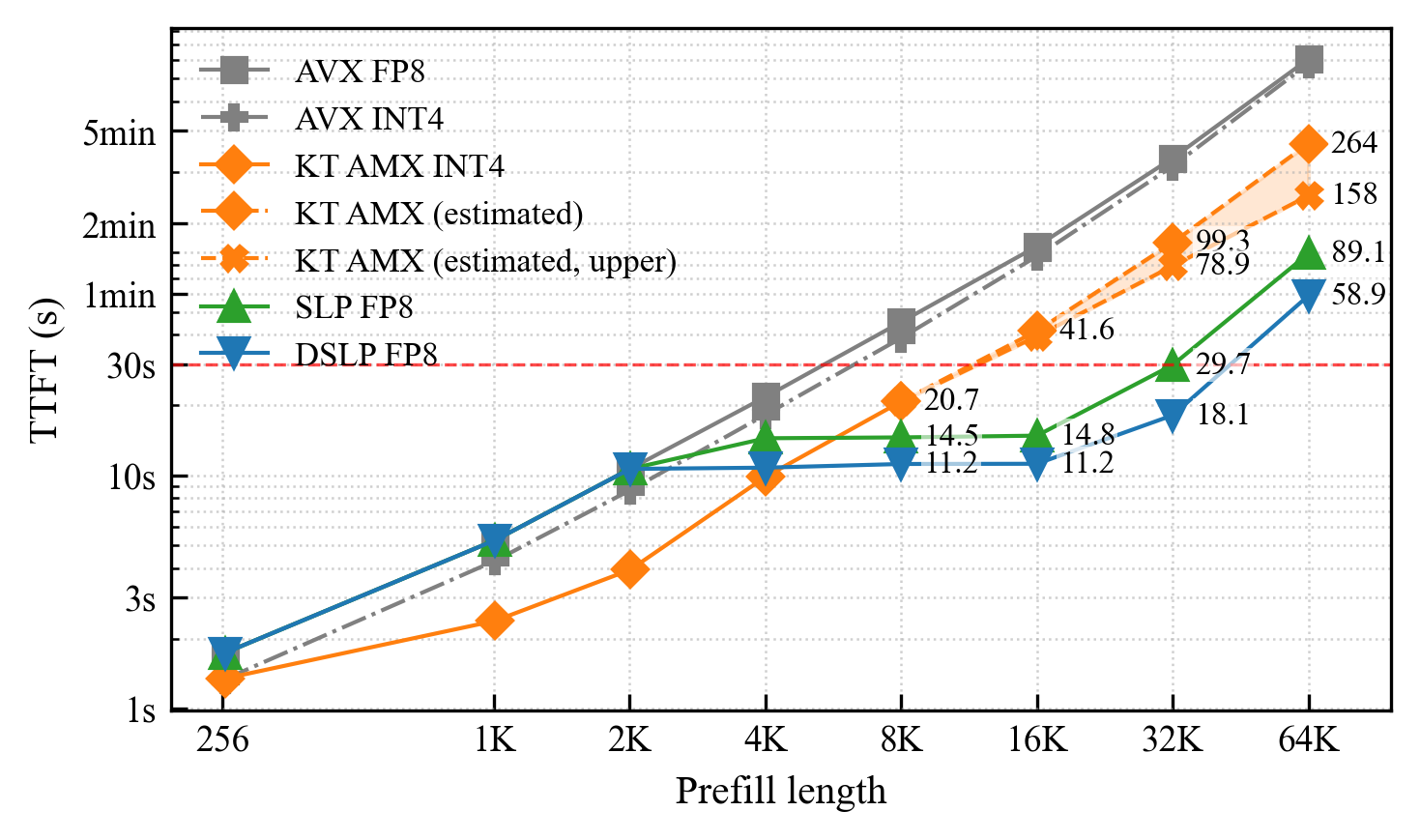}
    \caption{Time-to-First-Token (TTFT) comparison across prefill lengths from 256 to 64K tokens. We report CPU baselines using AVX FP8 and AVX INT4, KTransformers on AMX (KT AMX INT4), and our Stream‑Loading Prefill (SLP, FP8) and 2‑GPU Distributed SLP (DSLP, FP8). KT AMX results beyond 8K tokens are shown as estimates.}
    \label{fig:ttft}
\end{figure}

For a more comprehensive comparison, we evaluate TTFT across a wider range of prefill lengths. Because SLP and DSLP incur additional latency from weight transfer and stream setup, they are disadvantaged at very short prefixes. Consequently, we use the AVX FP8 kernel for requests up to 4K tokens and switch to SLP/DSLP for longer contexts. 

At small prefill lengths (256–2K), the CPU baselines are competitive: AVX FP8 and AVX INT4 track each other closely, and KT AMX INT4 achieves the lowest TTFT. Beyond 4K tokens, offloading with stream loading becomes clearly preferable. All CPU‑only configurations quickly exceed the 30~s TTFT threshold, while SLP and DSLP grow much more slowly. At 4K–32K tokens, SLP and DSLP keep TTFT below the 30~s target. For very long contexts (32–64K), SLP and especially DSLP remain much closer to practical latency budgets than KT‑AMX configuration. 

\subsection{Full-Precision Decode Performance}
\label{sec:eval:decode}

\begin{figure}[!h]
    \centering
    \includegraphics[width=0.9\linewidth]{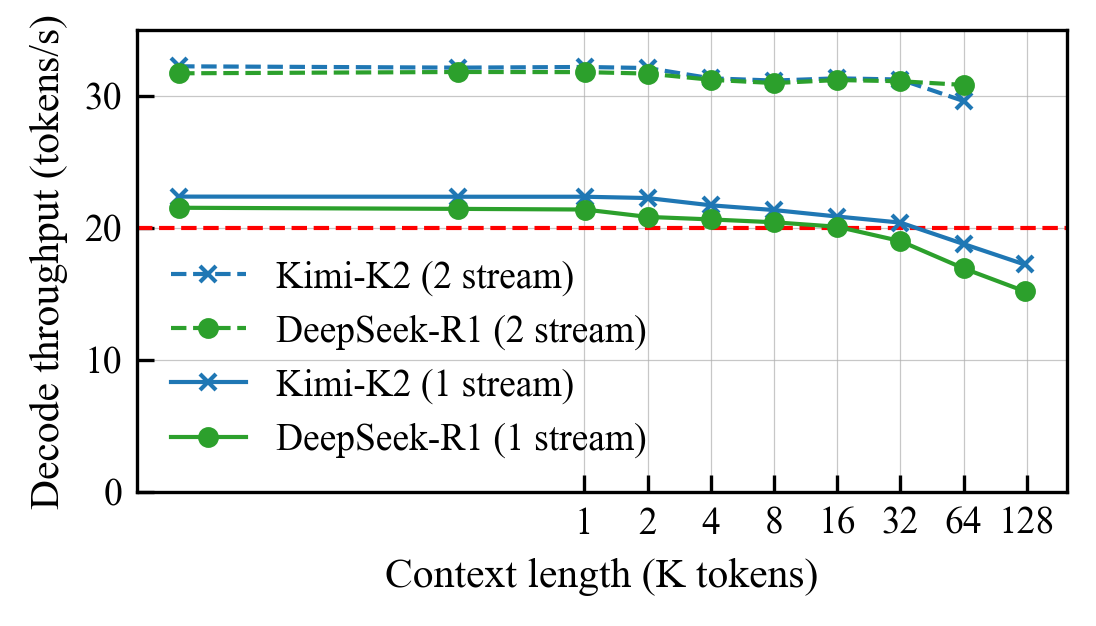}
    \caption{Decode throughput of the full-precision FP8 DeepSeek-R1 671B and Kimi-K2 1T model under single-stream and dual-stream execution across varying context lengths.}
    \label{fig:res-fp8}
\end{figure}

As shown in Figure~\ref{fig:res-fp8}, on our evaluation platform, the 671B-parameter DeepSeek-R1 model sustains 21.5 tokens/s on short prompts and maintains $\sim$20 tokens/s up to 32K context length. We observe similar performance with the Kimi-K2 model, which achieves comparable single-stream throughput of 22.4 tokens/s. With two equal-length, concurrent requests under our dual-batch parallel execution, total system throughput rises to 33.6 tokens/s and remains 31.1 tokens/s at 32K, ensuring each request still achieves $\sim$16 tokens/s. These results show that our design not only meets the single-stream QoS target but also scales gracefully under concurrency.

\begin{figure}
    \centering
    \includegraphics[width=0.95\linewidth]{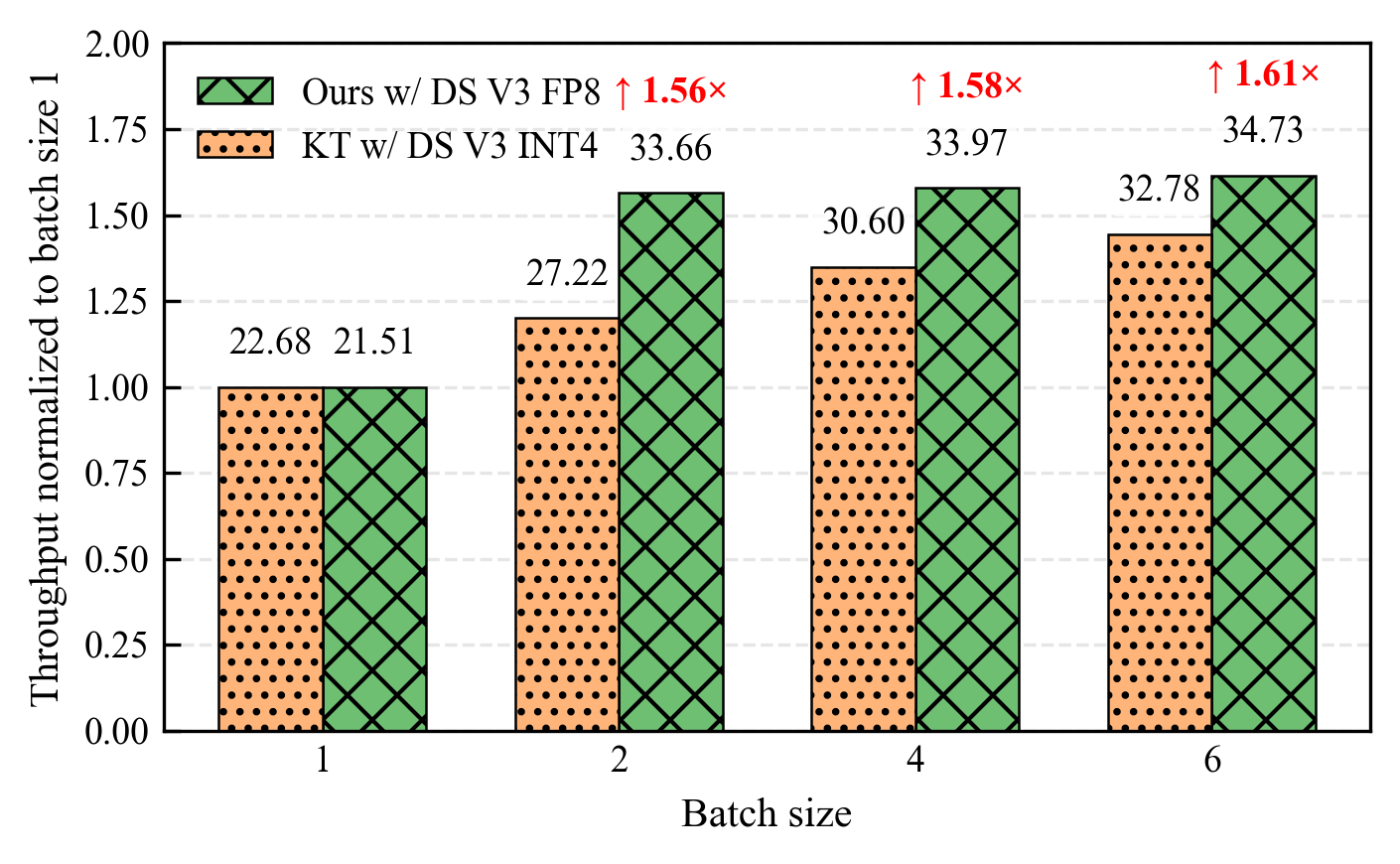}
    \caption{Decode throughput comparison of our engine running DeepSeek-R1 FP8 versus KTransformers (KT) running Q4\_K\_M (INT4) across batch sizes of 1, 2, 4, and 6. Bars are normalized to each engine’s batch size 1 result; absolute system throughputs (tokens/s) are annotated. Red markers report our throughput gains relative to batch size 1.}
    \label{fig:batch}
\end{figure}

Figure~\ref{fig:batch} compares decode throughput as a function of batch size for our engine running DeepSeek-R1 FP8 and KTransformers (KT) with Q4\_K\_M. Normalizing to batch size 1 highlights scaling efficiency: our engine achieves 1.56×, 1.58×, and 1.61× throughput at batch sizes 2, 4, and 6 respectively, while KT scales more modestly.

Per-request decode speed, computed as system throughput divided by batch size, shows our dual-batch Attention-MoE parallelism preserves user-level fluency under small batches: at batch size 2, our engine sustains 16.8 tok/s per request, a 21.7\% reduction from batch size 1, whereas KT drops from 22.7 to 13.6 tok/s (-40.0\%). This narrower per-request degradation indicates our design delivers better per-user responsiveness and higher aggregate throughput as concurrent requests increase.

\subsection{Quantized Model Decode Performance}
\label{sec:eval:decode_q4}

To contextualize our inference engine’s performance and validate our optimizations on mixed-precision inference, we also benchmarked the Q4\_K\_M variant of the DeepSeek-R1. While lower in numerical precision and sacrifices model fidelity, Q4\_K\_M quantized DeepSeek-R1 requires only 404GB of RAM to run, displaying its role in resource-constrained deployment scenarios. 

We conducted a head-to-head throughput comparison across varying context lengths from 1K to 128K tokens, evaluating: (1) Our engine, (2) llama.cpp, (3) ik\_llama.cpp, (4) KTransformers.

\begin{figure}
    \centering
    \includegraphics[width=0.9\linewidth]{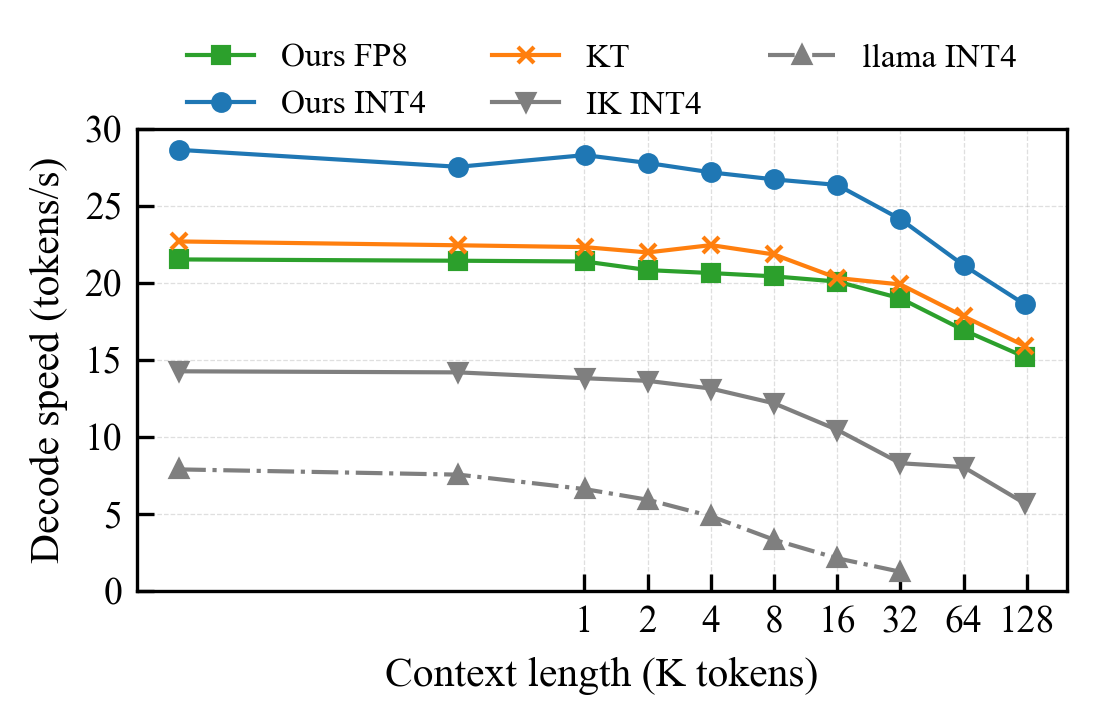}
    \caption{Decode speed comparison including Q4\_K\_M quantized DeepSeek-R1 671B model. IK: \texttt{ik\_llama.cpp}; llama: \texttt{llama.cpp}.}
    \label{fig:res-q4}
\end{figure}

As shown in Fig.~\ref{fig:res-q4}, our design consistently leads across all context lengths. At short sequences (1K–8K), it sustains 28 tokens/s, nearly doubling the throughput of ik\_llama.cpp (14 tokens/s) and outperforms KTransformers (22 tokens/s) by 25\%. As the context grows, competing engines degrade sharply: KTransformers drops to around 15 tokens/s at 128K. In contrast, our engine declines gracefully, maintaining 19 tokens/s at the maximum length. KTransformers on INT4 quantized model reaches similar decode speed to our engine running the original FP8 model. These results highlight the impact of our tailored CPU backend and parallelism design.

\subsection{Performance Under Concurrent Requests}
\label{sec:eval:concurrency}

\begin{figure*}[]
    \centering
    \includegraphics[width=\textwidth]{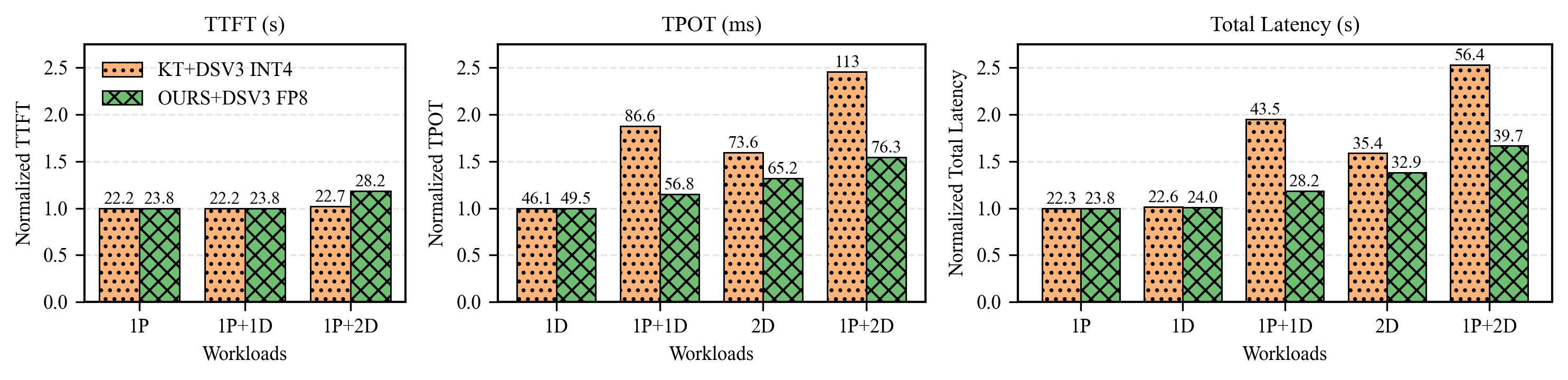}
    \caption{Microbenchmark of system latency under concurrent workloads. We compare KT with INT4 DeepSeek-R1 against our engine with FP8 DeepSeek-R1. All results are normalized to the first bar group in each subfigure, with the corresponding absolute values annotated on top of the bars. Workload labels: “1P” denotes one prefill request, while “1D” and “2D” denote one and two concurrent decode requests, respectively.}
    \label{fig:concurrency}
\end{figure*}

We evaluate concurrency in four representative scenarios to measure engine performance under concurrent prefill–decode and batched decode workloads. For isolation, we construct requests that each lasts about 20 seconds when running alone. Considering the different prefill throughput of our disaggregated SLP and KT on the AVX platform, we set the prefill length to 4K tokens for KT and 24K tokens for our engine, with no shared prefixes and only a few decoded tokens. Decode requests use short 10-token prompts and are forced to generate 480 tokens. Under these settings, when requests arrive simultaneously (regardless of whether they are prefill or decode), their execution could largely overlaps in time, allowing us to minimize variability and focus on end-to-end latency metrics.

Workloads consist of: (1) a single prefill request (1P), (2) a single decode request (1D) (these two serve as baselines), (3) one prefill request plus one decode request (1P+1D), (4) two concurrent decode requests (2D), and (5) one prefill request plus two decode requests (1P+2D). Each workload is executed multiple times on each engine. For metrics, we report the average TTFT for prefill requests and the average TPOT for decode requests, and we also measure the total latency to complete all requests within each workload.

We summarize the results in three perspectives: prefill latency (TTFT), decode latency (TPOT), and total workload latency, as shown in Figure~\ref{fig:concurrency}. Overall, our engine consistently handles concurrent workloads more robustly than KT. From the prefill perspective, TTFT remains stable for both engines under 1P and 1P+1D. For our engine, this is because prefill--decode disaggregation isolates the 1P request from concurrent decodes. Under 1P+2D, however, the additional DRAM traffic introduced by Attention--MoE parallelism slightly delays Stream-Loading Prefill's weight transfers, leading to an 18\% TTFT increase. We expect this effect to become slightly more pronounced as the number of concurrent decode requests further increases, because denser decode-side DRAM accesses leave less bandwidth headroom for SLP weight streaming. For KT, TTFT behavior is governed by its scheduling strategy of executing one 256-token chunked prefill per decode step, effectively prioritizing prefill over decode; this tradeoff becomes clearer when we examine TPOT.

From the decode perspective, our engine maintains stable TPOT across all workloads, while KT’s TPOT degrades significantly whenever prefill and decode coexist. KT’s average TPOT increases by up to 2.45× over the 1D baseline, whereas ours increases by at most 1.54× and scales smoothly under various concurrency workloads. 

Finally, in terms of total latency to complete all requests in each workload, our engine is consistently better than KT: our worst-case latency increase is 1.67× over the baseline, compared to KT’s 2.53×, and we also exhibit more favorable scaling when serving multiple concurrent decode requests (2D and 1P+2D).

\subsection{FP8 GEMV Kernel Performance}
\label{sec:eval:fp8kernel}

To isolate kernel performance from end-to-end system effects, we evaluate our FP8 GEMV kernels under real MoE workload and compare them against established CPU libraries. As shown in Table~\ref{tab:kernels}, our optimized kernel achieves a latency of 15.5~µs and bandwidth utilization of 947~GB/s — surpassing the our standard kernel (21.7~µs, 678~GB/s) and significantly outperforming OpenBLAS\cite{OpenBLAS}, and AOCL-BLAS\cite{AMD_AOCL} across FP32 and BF16 formats in latency while matching or exceeding bandwidth utilization. These results demonstrate that our kernels unlocks 4–5× latency reduction over existing CPU libraries with native FP8 support and delicate design.

\begin{table}[t]
\centering
\caption{Comparison of GEMV kernels on MoE workload. 
Our implementation is the only one supporting FP8 natively, while 
remaining competitive in bandwidth and significantly reducing latency.}
\label{tab:kernels}
\begin{tabular}{lccccc}
\toprule
\textbf{Library} & \textbf{Format} & \textbf{Latency ($\mu$s)} & \textbf{BW (GB/s)} \\
\midrule
OpenBLAS      & FP32       & 69.5   & 844.42 \\
AOCL-BLAS     & FP32       & 59.2   & 992.70 \\
AOCL-BLAS     & BF16       & 69.5   & 422.26 \\
\textbf{Ours (Std.)} & FP8 & 21.7    & 678.0 \\
\textbf{Ours (Optim.)} & FP8 & \textbf{15.5} & \textbf{947.1} \\
\bottomrule
\end{tabular}
\end{table}

\subsection{End-to-End Quality Evaluation}
\label{sec:eval:quality}

We evaluate end-to-end model quality on benchmark tasks. Table~\ref{tab:e2e-quality} reports Exact-Match results on MMLU-Redux~\cite{mmlu_redux} and MMLU-Pro~\cite{mmlu_pro} for our deployments of DeepSeek-V3.1~\cite{liu2024deepseek} and Kimi-K2-Instruct~\cite{moonshot_ai_2025}, compared against the officially published results of these models.

\begin{table}[ht]
\centering
\small
\caption{End-to-end quality evaluation using Exact Match. Official results are taken from the corresponding model releases.}
\label{tab:e2e-quality}
\begin{tabular}{@{}llcc@{}}
\toprule
\textbf{Model} & \textbf{Benchmark} & \textbf{Ours} & \textbf{Official} \\
\midrule
DeepSeek-V3.1 & MMLU-Redux & 91.00 & 91.8 \\
              & MMLU-Pro   & 83.41 & 83.7 \\
\addlinespace[0.3em]
Kimi-K2-Instruct & MMLU-Redux & 91.53 & 92.7 \\
                 & MMLU-Pro   & 80.00 & 81.1 \\
\bottomrule
\end{tabular}
\end{table}

Overall, our system preserves end-to-end quality well: the measured scores are close to the official references, with degradations within about 1--1.2 percentage points across the evaluated benchmarks. We believe these differences are minor and likely stem from small mismatches in generation configuration, rather than from a systematic loss of model capability introduced by our system. These results therefore provide additional evidence that our design preserves the quality of the original models while improving local serving performance.

\section{Related Work}

\paragraph{Cloud-service inference systems} GPU-centric engines such as vLLM, SGLang and TensorRT-LLM are optimized for multi-GPU nodes and datacenters to sustain high concurrency and throughput~\cite{kwon2023efficient, zheng2024sglang, tensorrt}. They converge on KV cache management (paged or radix-indexed)\cite{kwon2023efficient, zheng2024sglang}, concurrency-aware scheduling (Chunked Prefill, Continuous Batching)\cite{agrawal2023sarathi, yu2022orca}, distributed inference across context, sequence, tensor, pipeline, and expert parallelism~\cite{yang2024context, liu2023ring, jacobs2023ulysses, shoeybi2019megatron, zhu2024nanoflow, deepep2025, qin2025mooncake, lepikhin2020gshard}, and operator-level optimizations (kernel fusion, FlashAttention variants, new precisions)~\cite{dao2022flashattention, dao2023flashattention, shah2024flashattention, chitu}. While highly effective, these designs assume abundant GPU memory bandwidth, large aggregated VRAM, and fast interconnects, limiting flexibility and cost-efficiency outside GPU-first environments. Our work targets the local regime, adapting MoE-specific execution and parallelism to small-EP, low-concurrency settings on commodity CPU–GPU platforms while retaining cloud-level QoS.

\paragraph{Local inference systems for MoE models} Mainstream frameworks~\cite{zheng2024sglang, kwon2023efficient, tensorrt} prioritize datacenters and are impractical under constrained VRAM and bandwidth. Consumer-first solutions like Llama.cpp~\cite{llamacpp}, ik\_llama.cpp~\cite{ikawrakow_ik_llama_cpp}, and KTransformers~\cite{chen2025ktransformers} enable CPU–GPU hybrid serving; KTransformers combines quantization, rerouting, advanced CPU instructions, and NUMA-aware orchestration to deliver competitive throughput on commodity platforms. We advance this trajectory by achieving cloud-level SLOs for intact, original-precision MoE models on consumer CPU–GPU systems.
Compared with KTransformers, our work differs most clearly in target of achieving cloud-grade SLOs rather than enabling inference, and technically in three places: long-context prefill is made explicitly GPU-centric through stream loading instead of remaining CPU-centric; expert parallelism is redesigned for small EP sizes on PCIe-only local nodes; and concurrency is improved through intra-node prefill/decode disaggregation and dual-batch overlap rather than only through chunking and scheduling. 

\paragraph{Expert offloading techniques.}
Serial expert-processing approaches such as ES-MoE~\cite{es-moe} improve efficiency in memory-constrained settings by executing experts sequentially, avoiding zero-padding and large dispatch masks while improving load balance. However, ES-MoE differs from our target and design in three aspects. First, it targets training, where offloading and sequential processing enable larger microbatches and higher throughput, whereas our work targets latency-critical local inference, especially low-concurrency, long-context prefilling where TTFT is the primary objective. Second, ES-MoE plans expert transfer based on routing decisions, but this routing-dependent synchronization point leaves less opportunity to overlap transfer with the substantial pre-MoE computation on the critical path, which accounts for more than 40\% of per-layer execution time on a 20K-token prefill. In contrast, our SLP strategy stream-loads expert weights at a finer granularity and overlaps transfer with dense-module execution as well as expert computation. Third, for multi-GPU local prefilling, we further introduce SmallEP to reduce communication overhead under small EP sizes, improving distributed-prefill TTFT in PCIe-class local deployments.

\section{Discussion} \label{sec:discussion}

\paragraph{Lower-end CPUs} While prefill is largely addressed by our stream-loading prefill design, the main impact of lower-end CPUs is on decode. Small-batch decoding is well known to be heavily memory-bound. For the MoE computation executed on CPUs, the involved skinny GEMMs have an arithmetic intensity of roughly \(2\times\) batch size. In local deployment, the batch size is typically small (1--4), which remains far below the compute-to-memory ratio of current CPUs (around 50 FLOPs/byte). As a result, for lower-end CPU configurations, the dominant concern is reduced effective DRAM bandwidth rather than raw compute capability. According to our profiling, CPU-side MoE execution accounts for about 60\% of total decode time, with even greater exposure under dual-batch decode. We therefore expect decode performance to scale approximately linearly with available DRAM bandwidth, both in single-stream decoding and under dual-batch overlap.

\paragraph{Lower-VRAM or mid-range GPUs} The main hardware sensitivity on the GPU side lies in SLP/DSLP. For moderate prefill lengths (e.g., 16K tokens), performance depends on a tradeoff between experts ring-buffer capacity in VRAM and the ability to overlap weight transfer with computation. Our default design uses a full ring buffer covering all experts in one layer, which best smooths intra-layer transfer/computation imbalance and maximizes overlap. For DeepSeek-V3 and Kimi-K2, this corresponds to 11.3--16.9~GB of ring-buffer space, or 5.6--8.5~GB per GPU when distributed across two GPUs. If a lower-VRAM GPU cannot support this buffer size, overlap becomes less effective, and TTFT for medium-length prefills may degrade. By contrast, at very long prefills (e.g., beyond 50K tokens), the workload is dominated more by computation than by transfer, so performance should depend more directly on the GPU's available compute throughput.

\section{Conclusion}

This work demonstrates that flagship large MoE models can be served locally with cloud-level quality of service on commodity CPU–GPU platforms. By introducing stream-loading prefill (SLP) and its distributed variant (DSLP) with SmallEP, we achieve 30-second TTFT for 32K–45K prompts. Our intra-node prefill–decode disaggregation and dual-batch attention–MoE overlap sustains concurrency with minimal interference, while an AVX-512–optimized FP8 GEMV and fine-grained CPU parallelism close decode latency gaps for both full-precision and quantized models. Evaluations against mainstream CPU frameworks and BLAS libraries show that our design attains >20 TPS decode responsiveness and high-throughput prefills at substantially lower cost than GPU-centric deployments. Together, these results reshape the local MoE deployment landscape, enabling intact, original-precision inference and robust QoS without datacenter infrastructure, and point to a path for broader, cost-effective access to state-of-the-art applications.

\section*{Acknowledgments}

This work was supported in part by the Brain Science and Brain-like Intelligence Technology---National Science and Technology Major Project under Grant No. 2025ZD0215500/2025ZD0215502; the Tsinghua University Initiative Scientific Research Program under Grant No. 2022Z11ZRB002; the National Key Research and Development Program of China under Grant No. 2025YFB3003200; and the Jiangsu Provincial Science and Technology Program under Grant No. BE2023005-3.

% \clearpage

%-------------------------------------------------------------------------------
\bibliographystyle{plain}
\bibliography{main.bib}

\end{document}